\documentclass{mynature}

\usepackage{chapterbib}

\usepackage{graphics} 
\usepackage{times} 
\usepackage{mathptmx}

\usepackage[small,bf]{caption}

\typearea{12}

\renewcommand\refname{References}
\renewcommand{\thefootnote}{\fnsymbol{footnote}}

\usepackage{sectsty}

\begin{document}


\bibliographystyle{naturemag}

\onecolumn

\section*{\ \newline \Large Simulating the joint evolution of quasars, galaxies\vspace*{0.1cm}\newline
    and their large-scale distribution}

\noindent{\sffamily Volker~Springel$^{1}$, %
Simon~D.~M.~White$^{1}$, %
Adrian~Jenkins$^{2}$, %
Carlos~S.~Frenk$^{2}$, \newline%
Naoki~Yoshida$^{3}$, %
Liang~Gao$^{1}$, %
Julio~Navarro$^{4}$, %
Robert~Thacker$^{5}$, %
Darren~Croton$^{1}$, \newline%
John~Helly$^{2}$, %
John~A.~Peacock$^{6}$, %
Shaun~Cole$^{2}$, %
Peter~Thomas$^{7}$, %
Hugh~Couchman$^{5}$, \newline%
August~Evrard$^{8}$, %
J\"org~Colberg$^{9}$ \& %
Frazer~Pearce$^{10}$}
\ \\

\noindent%
{\footnotesize\it%
$^{1}${Max-Planck-Institute for Astrophysics, Karl-Schwarzschild-Str. 1, 85740 Garching, Germany}\\
$^{2}${Inst. for Computational Cosmology, Dep. of Physics, Univ. of
  Durham, South Road, Durham  DH1 3LE, UK}\\
$^{3}${Department of Physics, Nagoya University, Chikusa-ku, Nagoya
  464-8602, Japan}\\
$^{4}${Dep. of Physics \& Astron., University of
    Victoria, Victoria, BC, V8P 5C2, Canada}\\
$^{5}${Dep. of Physics \& Astron., McMaster Univ., 1280 Main
  St. West, Hamilton, Ontario, L8S 4M1, Canada}\\
$^{6}${Institute of Astronomy, University of Edinburgh, Blackford Hill, Edinburgh EH9 3HJ, UK}\\
$^{7}${Dep. of Physics \& Astron., University of
    Sussex, Falmer, Brighton BN1 9QH, UK}\\
$^{8}${Dep. of Physics \& Astron., Univ. of Michigan,
    Ann Arbor, MI 48109-1120, USA}\\
$^{9}${Dep. of Physics \& Astron., Univ. of Pittsburgh,
    3941 O'Hara Street, Pittsburgh PA 15260, USA}\\
$^{10}${Physics and Astronomy Department, Univ. of Nottingham,
    Nottingham NG7 2RD, UK}\\
}

\baselineskip26pt 
\setlength{\parskip}{12pt}
\setlength{\parindent}{22pt}%

\noindent{\bf The cold dark matter model has become the leading
  theoretical paradigm for the formation of structure in the Universe.
  Together with the theory of cosmic inflation, this model makes a
  clear prediction for the initial conditions for structure formation
  and predicts that structures grow hierarchically through
  gravitational instability.  Testing this model requires that the
  precise measurements delivered by galaxy surveys can be compared to
  robust and equally precise theoretical calculations. Here we present
  a novel framework for the quantitative physical interpretation of
  such surveys. This combines the largest simulation of the growth of
  dark matter structure ever carried out with new techniques for
  following the formation and evolution of the visible components.  We
  show that baryon-induced features in the initial conditions of the
  Universe are reflected in distorted form in the low-redshift galaxy
  distribution, an effect that can be used to constrain the nature of
  dark energy with next generation surveys.}

Recent large surveys such as the 2 degree Field Galaxy Redshift Survey
(2dFGRS) and the Sloan Digital Sky Survey (SDSS) have characterised
much more accurately than ever before not only the spatial clustering,
but also the physical properties of low-redshift galaxies.  Major
ongoing campaigns exploit the new generation of 8m-class telescopes
and the Hubble Space Telescope to acquire data of comparable quality
at high redshift. Other surveys target the weak image shear caused by
gravitational lensing to extract precise measurements of the
distribution of dark matter around galaxies and galaxy clusters. The
principal goals of all these surveys are to shed light on how galaxies
form, to test the current paradigm for the growth of cosmic structure,
and to search for signatures which may clarify the nature of dark
matter and dark energy.  These goals can be achieved only if the
accurate measurements delivered by the surveys can be compared to
robust and equally precise theoretical predictions.  Two problems have
so far precluded such predictions: (i) accurate estimates of
clustering require simulations of extreme dynamic range, encompassing
volumes large enough to contain representative populations of rare
objects (like rich galaxy clusters or quasars), yet resolving the
formation of individual low luminosity galaxies; (ii) critical aspects
of galaxy formation physics are uncertain and beyond the reach of
direct simulation (for example, the structure of the interstellar
medium, its consequences for star formation and for the generation of
galactic winds, the ejection and mixing of heavy elements, AGN feeding
and feedback effects \ldots) -- these must be treated by
phenomenological models whose form and parameters are adjusted by
trial and error as part of the overall data-modelling process. We have
developed a framework which combines very large computer simulations
of structure formation with post-hoc modelling of galaxy formation
physics to offer a practical solution to these two entwined problems.
  
During the past two decades, the cold dark matter (CDM) model,
augmented with a dark energy field (which may take the form of a
cosmological constant `$\Lambda$'), has developed into the standard
theoretical paradigm for galaxy formation.  It assumes that structure
grew from weak density fluctuations present in the otherwise
homogeneous and rapidly expanding early universe.  These fluctuations
are amplified by gravity, eventually turning into the rich structure
that we see around us today.  Confidence in the validity of this model
has been boosted by recent observations.  Measurements of the cosmic
microwave background (CMB) by the WMAP satellite\cite{Bennett2003}
were combined with the 2dFGRS to confirm the central tenets of the
model and to allow an accurate determination of the geometry and
matter content of the Universe about $380\,000$ years after the Big
Bang\cite{Spergel2003}.  The data suggest that the early density
fluctuations were a Gaussian random field, as predicted by
inflationary theory, and that the current energy density is dominated
by some form of dark energy. This analysis is supported by the
apparent acceleration of the current cosmic expansion inferred from
studies of distant supernovae\cite{Riess1998,Perlmutter1999}, as well
as by the low matter density derived from the baryon fraction of
clusters\cite{White1993}.

While the initial, linear growth of density perturbations can be
calculated analytically, the collapse of fluctuations and the
subsequent hierarchical build-up of structure is a highly nonlinear
process which is only accessible through direct numerical
simulation\cite{Davis1985}.  The dominant mass component, the cold
dark matter, is assumed to be made of elementary particles that
currently interact only gravitationally, so the collisionless dark
matter fluid can be represented by a set of discrete point
particles. This representation as an N-body system is a coarse
approximation whose fidelity improves as the number of particles in
the simulation increases.  The high-resolution simulation described
here -- dubbed the {\it Millennium Simulation} because of its size --
was carried out by the Virgo Consortium, a collaboration of British,
German, Canadian, and US astrophysicists. It follows $N= 2160^3\simeq
1.0078\times 10^{10}$ particles from redshift $z=127$ to the present
in a cubic region $500\,h^{-1}{\rm Mpc}$ on a side, where $1+z$ is the
expansion factor of the Universe relative to the present and $h$ is
Hubble's constant in units of $100\,{\rm km\,s^{-1}Mpc^{-1}}$. With
ten times as many particles as the previous largest computations of
this kind\cite{Colberg2000,Evrard2002,Wambsganss2004} (see
Supplementary Information), it offers substantially improved spatial
and time resolution within a large cosmological volume. Combining this
simulation with new techniques for following the formation and
evolution of galaxies, we predict the positions, velocities and
intrinsic properties of all galaxies brighter than the Small
Magellanic Cloud throughout volumes comparable to the largest current
surveys.  Crucially, this also allows us to establish evolutionary
links between objects observed at different epochs.  For example, we
demonstrate that galaxies with supermassive central black holes can
plausibly form early enough in the standard cold dark matter cosmology
to host the first known quasars, and that these end up at the centres
of rich galaxy clusters today.

\begin{figure*}
\noindent\hspace*{-0.5cm}%
\resizebox{17.0cm}{!}{\includegraphics{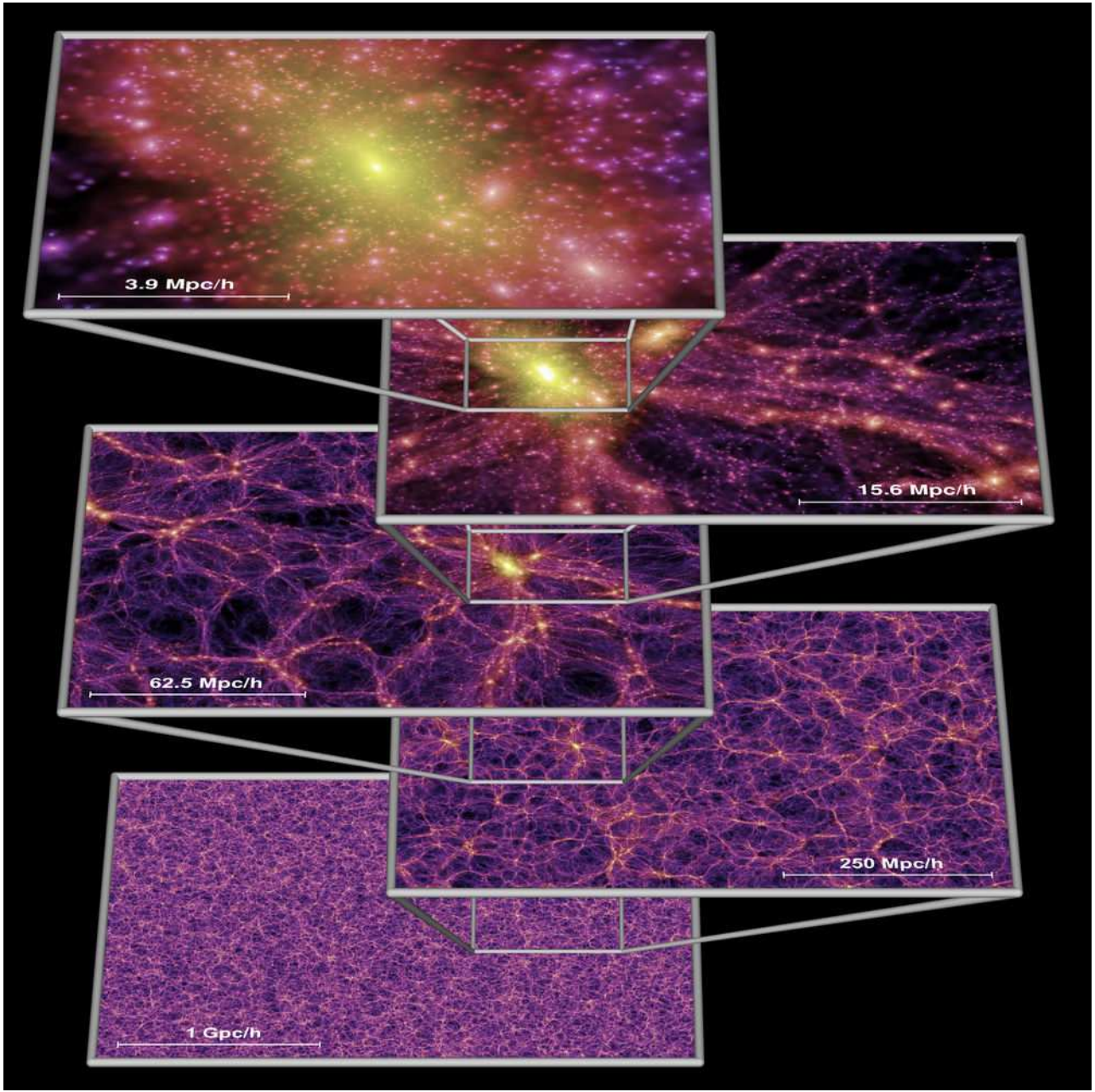}} %
\caption{The dark matter density field on various scales.  Each
  individual image shows the projected dark matter density field in a
  slab of thickness $15\,h^{-1}{\rm Mpc}$ (sliced from the periodic
  simulation volume at an angle chosen to avoid replicating structures
  in the lower two images), colour-coded by density and local dark
  matter velocity dispersion.  The zoom sequence displays consecutive
  enlargements by factors of four, centred on one of the many galaxy
  cluster halos present in the simulation.}
\label{FigDMDist}
\end{figure*}

\subsubsection*{Dark matter halos and galaxies}
The mass distribution in a $\Lambda$CDM universe has a complex
topology, often described as a ``cosmic web'' \cite{Bond1996}. This is
visible in full splendour in Fig.~\ref{FigDMDist} (see also the
corresponding Supplementary Video). The zoomed out panel at the bottom
of the figure reveals a tight network of cold dark matter clusters and
filaments of characteristic size $\sim 100\,h^{-1} {\rm Mpc}$. On
larger scales, there is little discernible structure and the
distribution appears homogeneous and isotropic.  Subsequent images
zoom in by factors of four onto the region surrounding one of the many
rich galaxy clusters.  The final image reveals several hundred dark
matter substructures, resolved as independent, gravitationally bound
objects orbiting within the cluster halo. These substructures are the
remnants of dark matter halos that fell into the cluster at earlier
times.

\begin{figure}
\hspace*{-1.0cm}%
\resizebox{16.0cm}{!}{\includegraphics{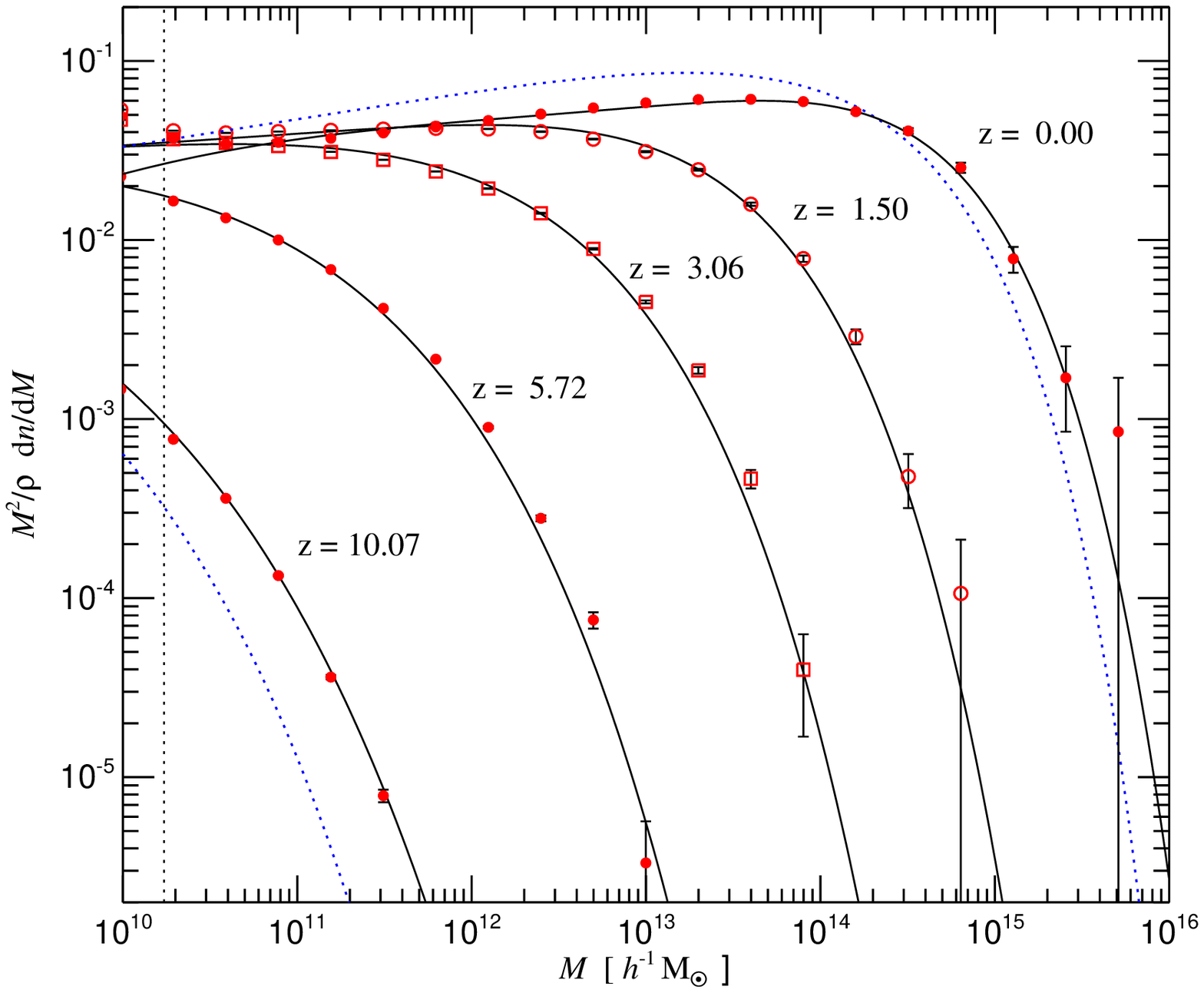}}
\caption{ \baselineskip20pt Differential halo number density as a
function of mass and epoch.  The function $n(M,z)$ gives the comoving
number density of halos less massive than $M$. We plot it as the halo
multiplicity function $M^2\rho^{-1}\,{\rm d}n/{\rm d}M$, where $\rho$
is the mean density of the universe. Groups of particles were found
using a friends-of-friends algorithm\cite{Davis1985} with linking
length equal to 0.2 of the mean particle separation. The fraction of
mass bound to halos of more than 20 particles (vertical dotted line)
grows from $6.42\times 10^{-4}$ at $z=10.07$ to 0.496 at $z=0$. Solid
lines are predictions from an analytic fitting function proposed in
previous work\cite{Jenkins2001}, while the dashed lines give the
Press-Schechter model\cite{Press1974} at $z=10.07$ and $z=0$.
\label{FigMassFunc} }
\end{figure}

The space density of dark matter halos at various epochs in the
simulation is shown in Fig.~\ref{FigMassFunc}. At the present day,
there are about 18 million halos above a detection threshold of 20
particles; 49.6\% of all particles are included in these halos.  These
statistics provide the most precise determination to date of the mass
function of cold dark matter halos\cite{Jenkins2001,Reed2003}. In the
range that is well sampled in our simulation ($z \le 12$, $M\ge
1.7\times 10^{10}\,h^{-1}{\rm M}_\odot$), our results are remarkably
well described by the analytic formula proposed by Jenkins et
al.\cite{Jenkins2001} from fits to previous simulations. Theoretical
models based on an ellipsoidal excursion set
formulation\cite{Sheth2002} give a less accurate, but still reasonable
match. However, the commonly used Press-Schechter
formula\cite{Press1974} underpredicts the high-mass end of the mass
function by up to an order of magnitude. Previous studies of the
abundance of rare objects, such as luminous quasars or clusters, based
on this formula may contain large errors\cite{Efstathiou1988}. We
return below to the important question of the abundance of quasars at
early times.

To track the formation of galaxies and quasars in the simulation, we
implement a semi-analytic model to follow gas, star and supermassive
black hole processes within the merger history trees of dark matter
halos and their substructures (see Supplementary Information).  The
trees contain a total of about 800 million nodes, each corresponding
to a dark matter subhalo and its associated galaxies. This methodology
allows us to test, during postprocessing, many different
phenomenological treatments of gas cooling, star formation, AGN
growth, feedback, chemical enrichment, etc. Here, we use an update of
models described in\cite{Springel2001b,Kauffmann2000} which are
similar in spirit to previous semi-analytic
models\cite{WhiteFrenk1991,Kauffmann1993,Cole1994,Baugh1996,Sommerville1999,Kauffmann1999};
the modelling assumptions and parameters are adjusted by trial and
error in order to fit the observed properties of low redshift
galaxies, primarily their joint luminosity-colour distribution and
their distributions of morphology, gas content and central black hole
mass.  Our use of a high-resolution simulation, particularly our
ability to track the evolution of dark matter substructures, removes
much of the uncertainty of the more traditional semi-analytic
approaches based on Monte-Carlo realizations of merger trees.  Our
technique provides accurate positions and peculiar velocities for all
the model galaxies.  It also enables us to follow the evolutionary
history of individual objects and thus to investigate the relationship
between populations seen at different epochs. It is the ability to
establish such evolutionary connections that makes this kind of
modelling so powerful for interpreting observational data.

\subsubsection*{The fate of the first quasars}

Quasars are among the most luminous objects in the Universe and can be
detected at huge cosmological distances.  Their luminosity is thought
to be powered by accretion onto a central, supermassive black
hole. Bright quasars have now been discovered as far back as redshift
$z=6.43$ (ref.~\cite{Fan2003}), and are believed to harbour central
black holes of mass a billion times that of the sun. At redshift
$z\sim 6$, their comoving space density is estimated to be $\sim (2.2
\pm 0.73)\times 10^{-9}\,h^3{\rm Mpc}^{-3}$ (ref.~\cite{Fan2004}).
Whether such extreme rare objects can form at all in a $\Lambda$CDM
cosmology is an open question.

A volume the size of the Millennium Simulation should contain, on
average, just under one quasar at the above space density.  Just what
sort of object should be associated with these ``first quasars'' is,
however, a matter of debate. In the local universe, it appears that
every bright galaxy hosts a supermassive black hole and there is a
remarkably good correlation between the mass of the central black hole
and the stellar mass or velocity dispersion of the bulge of the host
galaxy\cite{Tremaine2002}.  It would therefore seem natural to assume
that at any epoch, the brightest quasars are always hosted by the
largest galaxies. In our simulation, `large galaxies' can be
identified in various ways, for example, according to their dark
matter halo mass, stellar mass, or instantaneous star formation rate.
We have identified the 10 `largest' objects defined in these three
ways at redshift $z=6.2$. It turns out that these criteria all select
essentially the same objects: the 8 largest galaxies by halo mass are
identical to the 8 largest galaxies by stellar mass, only the ranking
differs.  Somewhat larger differences are present when galaxies are
selected by star formation rate, but the 4 first-ranked galaxies are
still amongst the 8 identified according to the other 2 criteria.

\begin{figure}
\vspace*{-1.0cm}\hspace*{-0.3cm}%
\resizebox{8.2cm}{!}{\includegraphics{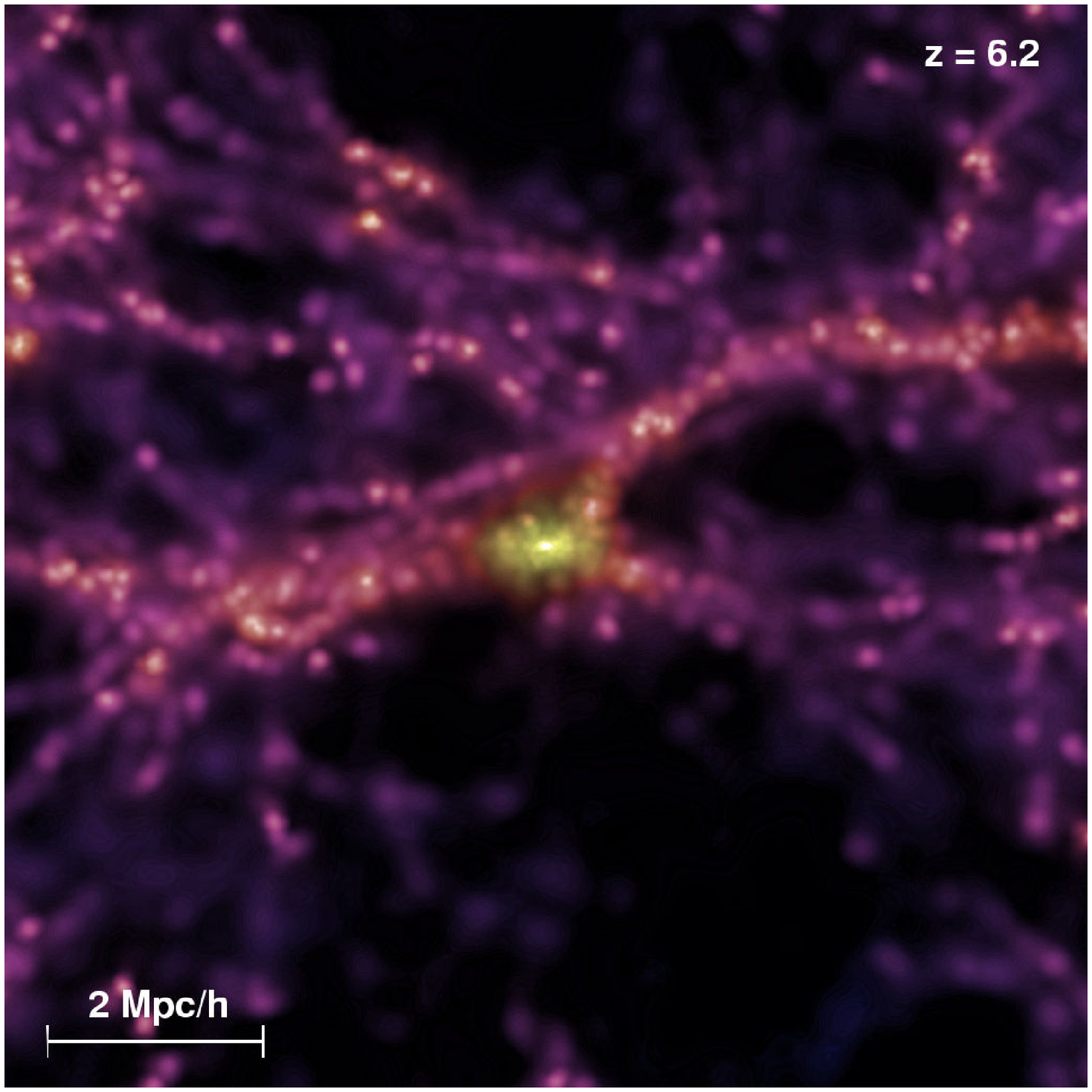}} %
\resizebox{8.2cm}{!}{\includegraphics{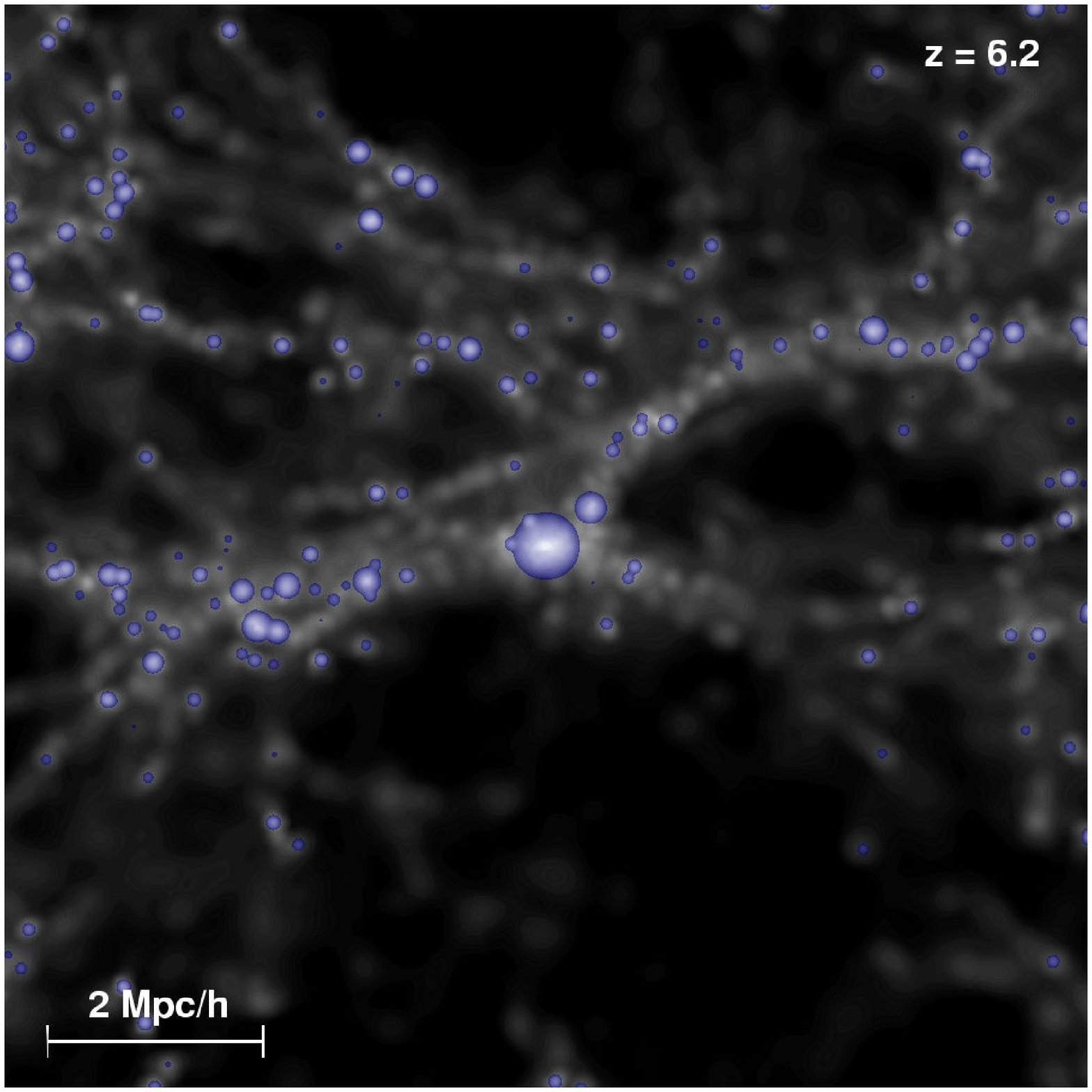}}\vspace*{0.05cm}\\%
\hspace*{-0.3cm}%
\resizebox{8.2cm}{!}{\includegraphics{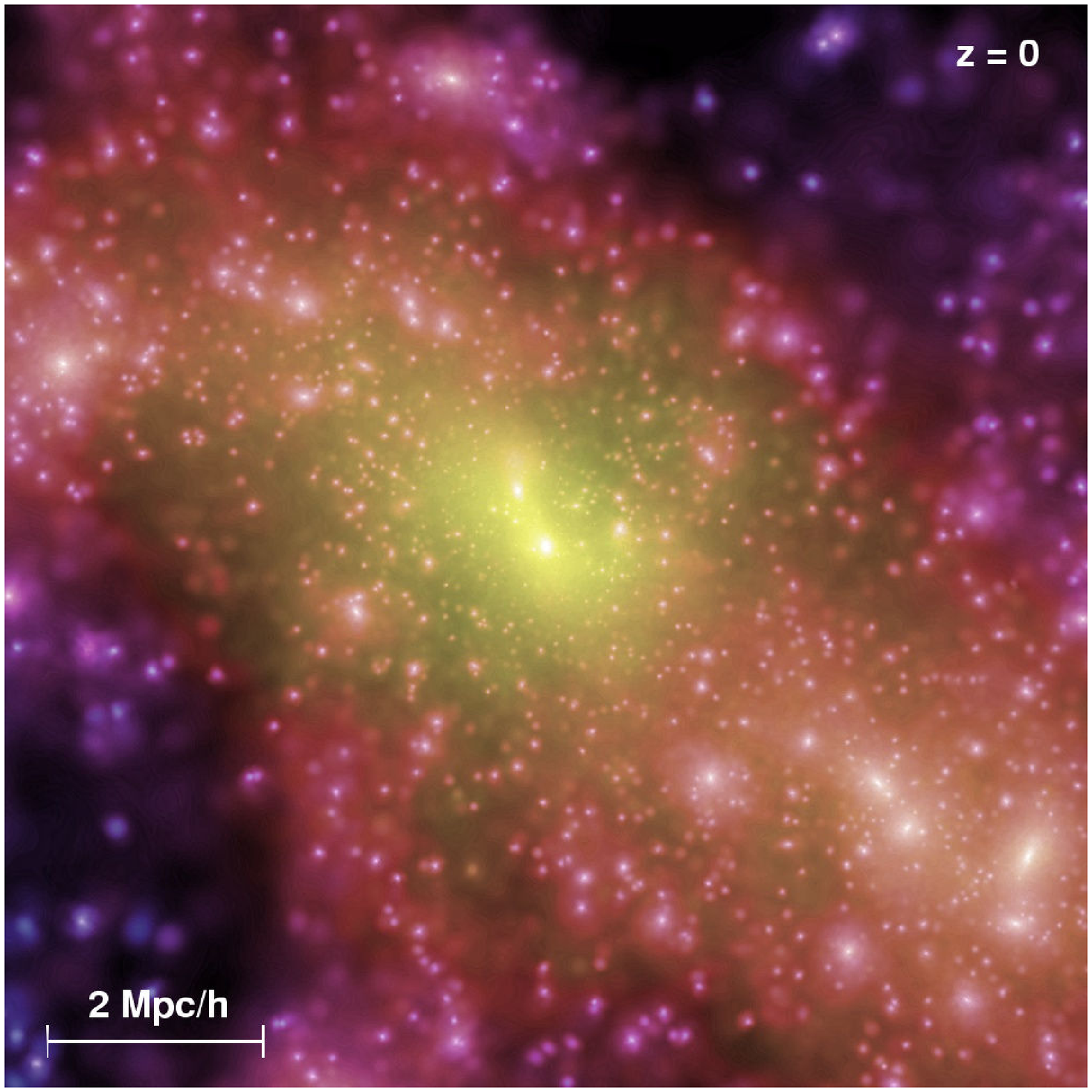}} %
\resizebox{8.2cm}{!}{\includegraphics{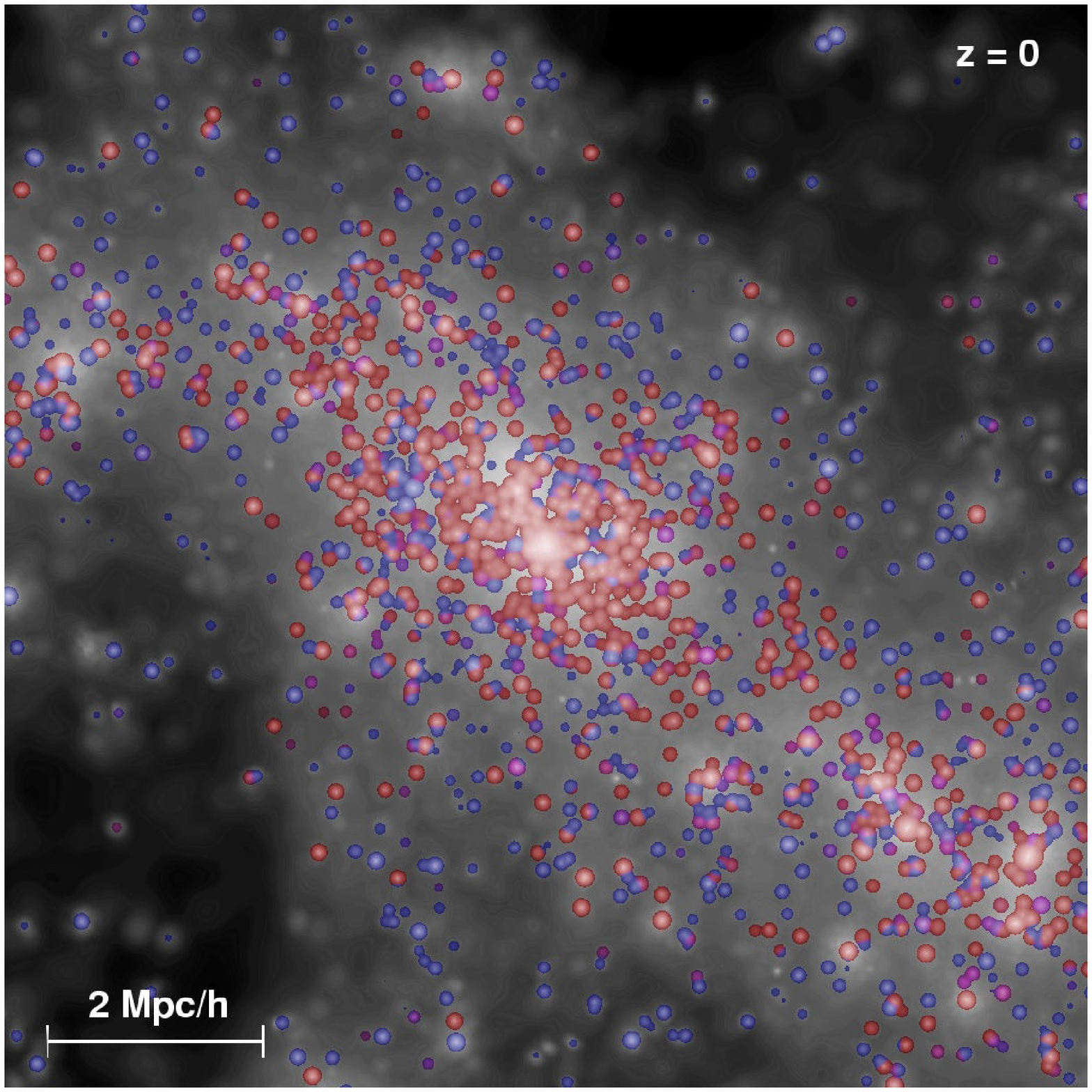}}\\%
\caption{Environment of a `first quasar candidate' at high and low redshifts.
  The two panels on the left show the projected dark matter
  distribution in a cube of comoving sidelength $10\,h^{-1}{\rm Mpc}$,
  colour-coded according to density and local dark matter velocity
  dispersion.  The panels on the right show the galaxies of the
  semi-analytic model overlayed on a gray-scale image of the dark
  matter density. The volume of the sphere representing each galaxy is
  proportional to its stellar mass, and the chosen colours encode the
  restframe stellar $B-V$ colour index. While at $z=6.2$ (top) all
  galaxies appear blue due to ongoing star formation, many of the
  galaxies that have fallen into the rich cluster at $z=0$ (bottom)
  have turned red.
\label{FigFirstQuasar}}
\end{figure}

In Figure~\ref{FigFirstQuasar}, we illustrate the environment of a
``first quasar'' candidate in our simulation at $z=6.2$.  The object
lies on one of the most prominent dark matter filaments and is
surrounded by a large number of other, much fainter galaxies. It has a
stellar mass of $6.8\times 10^{10}\,h^{-1}{\rm M}_\odot$, the largest
in the entire simulation at $z=6.2$, a dark matter virial mass of
$3.9\times 10^{12}\,h^{-1}{\rm M}_\odot$, and a star formation rate of
$235\, {\rm M_\odot yr^{-1}}$.  In the local universe central black
hole masses are typically $\sim 1/1000$ of the bulge stellar
mass\cite{Merrit2001}, but in the model we test here these massive
early galaxies have black hole masses in the range $10^8 - 10^9{\rm
M}_\odot$, significantly larger than low redshift galaxies of similar
stellar mass. To attain the observed luminosities, they must convert
infalling mass to radiated energy with a somewhat higher efficiency
than the $\sim 0.1\,c^2$ expected for accretion onto a {\em
non-spinning} black hole.

Within our simulation we can readily address fundamental questions
such as: ``Where are the descendants of the early quasars today?'', or
``What were their progenitors?''. By tracking the merging history
trees of the host halos, we find that all our quasar candidates end up
today as central galaxies in rich clusters. For example, the object
depicted in Fig.~\ref{FigFirstQuasar} lies, today, at the centre of
the ninth most massive cluster in the volume, of mass
$M=1.46\times10^{15}\,h^{-1}{\rm M}_\odot$. The candidate with the
largest virial mass at $z=6.2$ (which has stellar mass $4.7\times
10^{10}\,h^{-1}{\rm M}_\odot$, virial mass $4.85\times
10^{12}\,h^{-1}{\rm M}_\odot$, and star formation rate $218\, {\rm
M_\odot yr^{-1}}$) ends up in the second most massive cluster, of mass
$3.39\times10^{15}\,h^{-1}{\rm M}_\odot$.  Following the merging tree
backwards in time, we can trace our quasar candidate back to redshift
$z=16.7$, when its host halo had a mass of only $1.8\times
10^{10}\,h^{-1}{\rm M}_\odot$. At this epoch, it is one of just 18
objects that we identify as collapsed systems with $\ge 20$
particles. These results confirm the view that rich galaxy clusters
are rather special places.  Not only are they the largest virialised
structures today, they also lie in the regions where the first
structures developed at high redshift. Thus, the best place to search
for the oldest stars in the Universe or for the descendants of the
first supermassive black holes is at the centres of present-day rich
galaxy clusters.

\subsubsection*{The clustering evolution of dark matter and galaxies} 

\begin{figure}
\vspace*{-0.0cm}\ \\%
\resizebox{14.0cm}{!}{\includegraphics{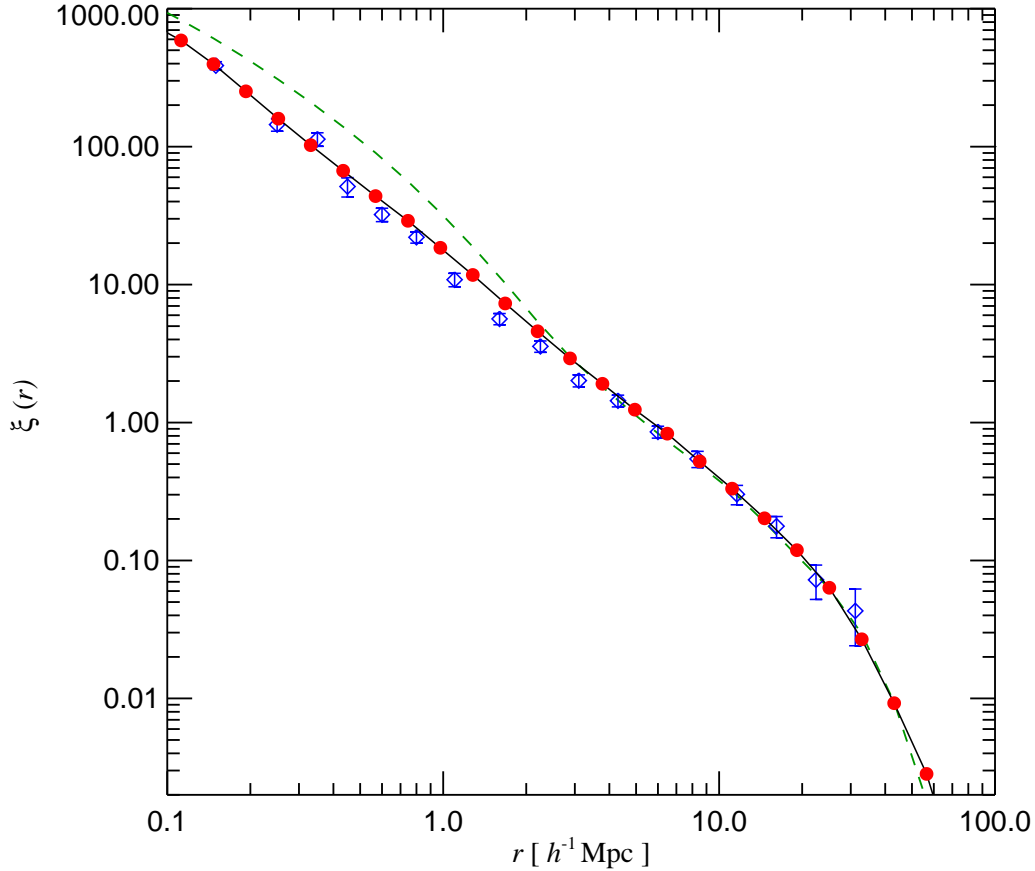}}\\
\caption{Galaxy 2-point correlation function at the present epoch.
  Red symbols (with vanishingly small Poisson error-bars) show
  measurements for model galaxies brighter than $M_K = -23$.  Data for
  the large spectroscopic redshift survey 2dFGRS\cite{Hawkins2003} are
  shown as blue diamonds.  The SDSS\cite{Zehavi2002} and
  APM\cite{Padilla2003} surveys give similar results. Both, for the
  observational data and for the simulated galaxies, the correlation
  function is very close to a power-law for $r\le 20\, h^{-1}{\rm
  Mpc}$. By contrast the correlation function for the dark matter
  (dashed line) deviates strongly from a power-law.
\label{FigClustering}}
\end{figure}

The combination of a large-volume, high-resolution N-body simulation
with realistic modelling of galaxies enables us to make precise
theoretical predictions for the clustering of galaxies as a function
of redshift and intrinsic galaxy properties. These can be compared
directly with existing and planned surveys.  The 2-point correlation
function of our model galaxies at redshift $z=0$ is plotted in
Fig.~\ref{FigClustering} and is compared with a recent measurement
from the 2dFGRS\cite{Hawkins2003}.  The prediction is remarkably close
to a power-law, confirming with much higher precision the results of
earlier semi-analytic\cite{Kauffmann1999,Benson2000} and
hydrodynamic\cite{Weinberg2004} simulations. This precision will allow
interpretation of the small, but measurable deviations from a pure
power-law found in the most recent data\cite{Padilla2003,Zehavi2004}.
The simple power-law form contrasts with the more complex behaviour
exhibited by the dark matter correlation function but is really no
more than a coincidence.  Correlation functions for galaxy samples
with different selection criteria or at different redshifts do not, in
general, follow power-laws.

\begin{figure}
\hspace*{-1.0cm}\ \resizebox{17cm}{!}{\includegraphics{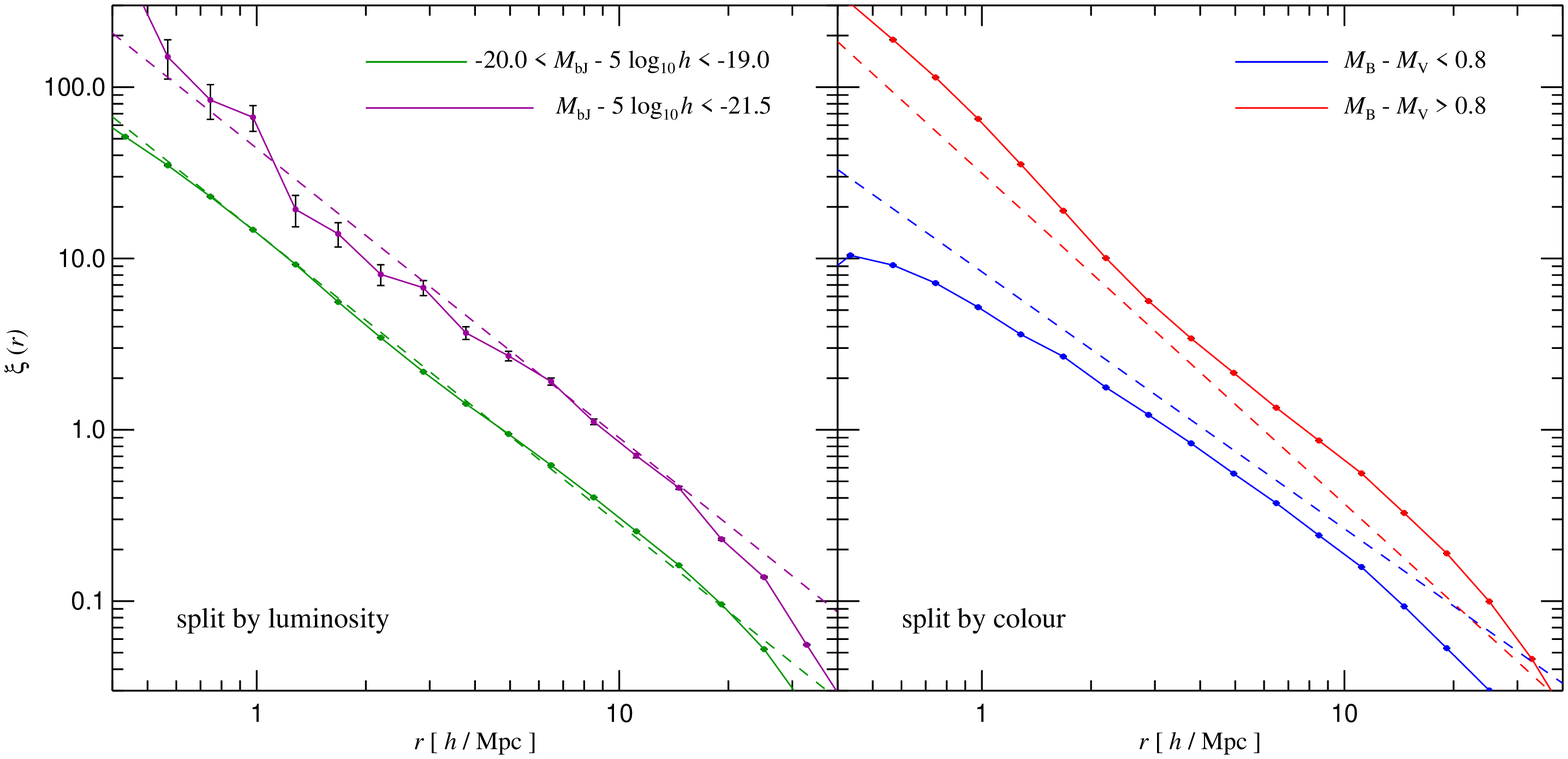}}
\caption{Galaxy clustering as a function of luminosity and colour.  In
  the panel on the left, we show the 2-point correlation function of
  our galaxy catalogue at $z=0$ split by luminosity in the bJ-band
  (symbols). Brighter galaxies are more strongly clustered, in
  quantitative agreement with observations\cite{Norberg2001} (dashed
  lines).  Splitting galaxies according to colour (right panel), we
  find that red galaxies are more strongly clustered with a steeper
  correlation slope than blue
  galaxies. Observations\cite{Madgwick2003} (dashed lines) show a
  similar trend, although the difference in clustering amplitude is
  smaller than in this particular semi-analytic model.
\label{FigClusteringSubsamples}}
\end{figure}

Although our semi-analytic model was not tuned to match observations
of galaxy clustering, in not only produces the excellent overall
agreement shown in Fig.~\ref{FigClustering}, but also reproduces the
observed dependence of clustering on magnitude and colour in the
2dFGRS and SDSS\cite{Norberg2001,Zehavi2002,Madgwick2003}, as shown in
Figure~\ref{FigClusteringSubsamples}.  The agreement is particularly
good for the dependence of clustering on luminosity. The colour
dependence of the slope is matched precisely, but the amplitude
difference is greater in our model than is
observed\cite{Madgwick2003}.  Note that our predictions for galaxy
correlations split by colour deviate substantially from
power-laws. Such predictions can be easily tested against survey data
in order to clarify the physical processes responsible for the
observed difference.

In contrast to the near power-law behaviour of galaxy correlations on
small scales, the large-scale clustering pattern may show interesting
structure.  Coherent oscillations in the primordial plasma give rise
to the well-known acoustic peaks in the
CMB\cite{deBernardis2000,Mauskopf2000,Spergel2003} and also leave an
imprint in the linear power spectrum of the dark matter. Detection of
these ``baryon wiggles'' would not only provide a beautiful
consistency check for the cosmological paradigm, but could also have
important practical applications. The characteristic scale of the
wiggles provides a ``standard ruler'' which may be used to constrain
the equation of state of the dark energy\cite{Blake2003}. A critical
question when designing future surveys is whether these baryon wiggles
are present and are detectable in the {\em galaxy} distribution,
particularly at high redshift.

On large scales and at early times, the mode amplitudes of the {\it
dark matter} power spectrum grow linearly, roughly in proportion to
the cosmological expansion factor. Nonlinear evolution accelerates the
growth on small scales when the dimensionless power $\Delta^2(k) = k^3
P(k)/(2\pi^2)$ approaches unity; this regime can only be studied
accurately using numerical simulations. In the Millennium Simulation,
we are able to determine the nonlinear power spectrum over a larger
range of scales than was possible in earlier work\cite{Jenkins1998},
almost five orders of magnitude in wavenumber $k$.

\begin{figure}
\begin{center}
\vspace*{-1.6cm}\hspace*{-1.0cm}%
\resizebox{15.8cm}{!}{\includegraphics{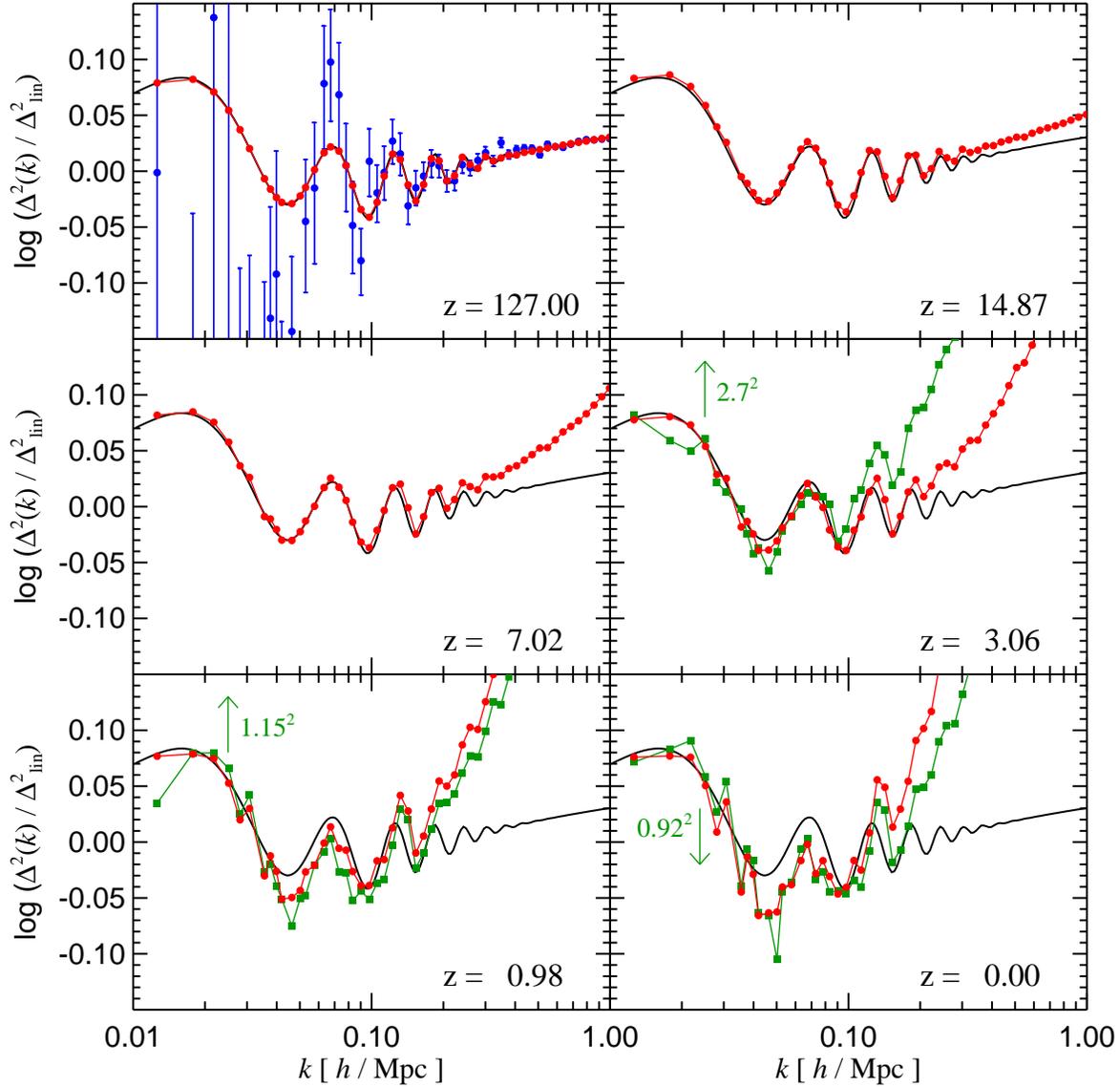}}\vspace*{-1.0cm}%
\end{center}
\caption{ Power spectra of the dark matter and galaxy distributions in
  the baryon oscillation region. All measurements have been divided by
  a linearly evolved, CDM-only power spectrum\cite{Bardeen1986}. Red
  circles show the dark matter, and green squares the galaxies. Blue
  symbols give the actual realization of the initial fluctuations in
  our simulation, which scatters around the mean input power (black
  lines) due to the finite number of modes.  Since linear growth
  preserves relative mode amplitudes, we correct the power in each bin
  to the expected input power and apply these scaling factors at all
  other times.  At $z=3.06$, galaxies with stellar mass above
  $5.83\times 10^9\,h^{-1}{\rm M}_\odot$ and space-density of $8\times
  10^{-3}\,h^{3}{\rm Mpc}^{-3}$ were selected. Their large-scale
  density field is biased by a factor $b=2.7$ with respect to the dark
  matter (the galaxy measurement has been divided by $b^2$). At $z=0$,
  galaxies brighter than $M_B = -17$ and a space density higher by a
  factor $\sim 7.2$ were selected. They exhibit a slight antibias,
  $b=0.92$.  Corresponding numbers for $z=0.98$ are $M_B = -19$ and
  $b=1.15$.
\label{FigWiggles}}
\end{figure}

At the present day, the acoustic oscillations in the matter power
spectrum are expected to fall in the transition region between linear
and nonlinear scales.  In Fig.~\ref{FigWiggles}, we examine the matter
power spectrum in our simulation in the region of the
oscillations. Dividing by the smooth power spectrum of a $\Lambda$CDM
model with no baryons\cite{Bardeen1986} highlights the baryonic
features in the initial power spectrum of the simulation, although
there is substantial scatter due to the small number of large-scale
modes.  Since linear growth preserves the relative mode amplitudes, we
can approximately correct for this scatter by scaling the measured
power in each bin by a multiplicative factor based on the initial
difference between the actual bin power and the mean power expected in
our $\Lambda$CDM model.  This makes the effects of nonlinear evolution
on the baryon oscillations more clearly visible.  As
Fig.~\ref{FigWiggles} shows, nonlinear evolution not only accelerates
growth but also reduces the baryon oscillations: scales near peaks
grow slightly more slowly than scales near troughs. This is a
consequence of the mode-mode coupling characteristic of nonlinear
growth. In spite of these effects, the first two ``acoustic peaks''
(at $k\sim 0.07$ and $k\sim 0.13\,h\,{\rm Mpc}^{-1}$, respectively) in
the dark matter distribution do survive in distorted form until the
present day (see the lower right panel of Fig.~\ref{FigWiggles}).

Are the baryon wiggles also present in the galaxy distribution?
Fig.~\ref{FigWiggles} shows that the answer to this important question
is `yes'. The $z=0$ panel shows the power spectrum for all model
galaxies brighter than $M_B = -17$. On the largest scales, the galaxy
power spectrum has the same shape as that of the dark matter, but with
slightly lower amplitude corresponding to an ``antibias'' of
8\%. Samples of brighter galaxies show less antibias while for the
brightest galaxies, the bias becomes slightly positive. The figure
also shows measurements of the power spectrum of luminous galaxies at
redshifts $z=0.98$ and $z=3.06$. Galaxies at $z=0.98$ were selected to
have a magnitude $M_B<-19$ in the restframe, whereas galaxies at
$z=3.06$ were selected to have stellar mass larger than $5.83\times
10^9\,h^{-1}{\rm M}_\odot$, corresponding to a space density of
$8\times 10^{-3}\,h^{3}{\rm Mpc}^{-3}$, similar to that of the
Lyman-break galaxies observed at $z\sim 3$\cite{Adelberger1998}.
Signatures of the first two acoustic peaks are clearly visible at both
redshifts, even though the density field of the $z=3$ galaxies is much
more strongly biased with respect to the dark matter (by a factor
$b=2.7$) than at low redshift.  Selecting galaxies by their star
formation rate rather than their stellar mass (above $10.6\,{\rm
M_\odot yr^{-1}}$ for an equal space density at $z=3$) produces very
similar results.

Our analysis demonstrates conclusively that baryon wiggles should
indeed be present in the galaxy distribution out to redshift
$z=3$. This has been assumed but not justified in recent proposals to
use evolution of the large-scale galaxy distribution to constrain the
nature of the dark energy.  To establish whether the baryon
oscillations can be measured in practice with the requisite accuracy
will require detailed modelling of the selection criteria of an actual
survey and a thorough understanding of the systematic effects that
will inevitably be present in real data. These issues can only be
properly addressed by means of specially designed mock catalogues
constructed from realistic simulations. We plan to construct suitable
mock catalogues from the Millennium Simulation and make them publicly
available. Our provisional conclusion, however, is that the next
generation of galaxy surveys offers excellent prospects for
constraining the equation of state of the dark energy.

N-body simulations of CDM universes are now of such size and quality
that realistic modelling of galaxy formation in volumes matched to
modern surveys has become possible. Detailed studies of galaxy and AGN
evolution exploiting the unique dataset of the Millennium Simulation
therefore enable stringent new tests of the theory of hierarchical
galaxy formation.  Using the simulation we demonstrated that quasars
can plausibly form sufficiently early in a $\Lambda$CDM universe to be
compatible with observation, that their progenitors were already
massive by $z \sim 16$, and that their $z=0$ descendents lie at the
centres of cD galaxies in rich galaxy clusters. Interesting tests of
our predictions will become possible if observations of the black hole
demographics can be extended to high redshift, allowing, for example,
a measurement of the evolution of the relationship between
supermassive black hole masses and the velocity dispersion of their
host stellar bulges.

We have also demonstrated that a power-law galaxy autocorrelation
function can arise naturally in a $\Lambda$CDM universe, but that this
suggestively simple behaviour is merely a coincidence. Galaxy surveys
will soon reach sufficient statistical power to measure precise
deviations from power-laws for galaxy subsamples, and we expect that
comparisons of the kind we have illustrated will lead to tight
constraints on the physical processes included in the galaxy formation
modelling.  Finally, we have demonstrated for the first time that the
baryon-induced oscillations recently detected in the CMB power
spectrum should survive in distorted form not only in the nonlinear
dark matter power spectrum at low redshift, but also in the power
spectra of realistically selected galaxy samples at $0<z<3$. Present
galaxy surveys are marginally able to detect the baryonic features at
low redshifts\cite{Cole2005,Eisenstein2005}. If future surveys improve
on this and reach sufficient volume and galaxy density also at high
redshift, then precision measurements of galaxy clustering will shed
light on one of the most puzzling components of the universe, the
elusive dark energy field.

\vspace*{1cm}

\section*{Methods}

The Millennium Simulation was carried out with a specially customised
version of the {\small GADGET2}~(Ref. \cite{Springel2001})~code, using
the ``TreePM'' method\cite{Xu1995} for evaluating gravitational
forces. This is a combination of a hierarchical multipole expansion,
or ``tree'' algorithm\cite{Barnes1986}, and a classical, Fourier
transform particle-mesh method\cite{Hockney1981}. The calculation was
performed on 512 processors of an IBM p690 parallel computer at the
Computing Centre of the Max-Planck Society in Garching, Germany. It
utilised almost all the 1~TB of physically distributed memory
available.  It required about $350\,000$ processor hours of CPU time,
or 28 days of wall-clock time.  The mean sustained floating point
performance (as measured by hardware counters) was about 0.2~TFlops,
so the total number of floating point operations carried out was of
order $5\times 10^{17}$.

The cosmological parameters of our $\Lambda$CDM-simulation are:
$\Omega_{\rm m}= \Omega_{\rm dm}+\Omega_{\rm b}=0.25$, $\Omega_{\rm
b}=0.045$, $h=0.73$, $\Omega_\Lambda=0.75$, $n=1$, and
$\sigma_8=0.9$. Here $\Omega_{\rm m}$ denotes the total matter density
in units of the critical density for closure, $\rho_{\rm crit}=3
H_0^2/(8\pi G)$. Similarly, $\Omega_{\rm b}$ and $\Omega_\Lambda$
denote the densities of baryons and dark energy at the present
day. The Hubble constant is parameterised as $H_0 = 100\, h\, {\rm
km\, s^{-1} Mpc^{-1}}$, while $\sigma_8$ is the {\em rms} linear mass
fluctuation within a sphere of radius $8\, h^{-1}{\rm Mpc}$
extrapolated to $z=0$.  Our adopted parameter values are consistent
with a combined analysis of the 2dFGRS\cite{Colless2001} and first
year WMAP data\cite{Spergel2003}.

The simulation volume is a periodic box of size $500\,h^{-1}{\rm Mpc}$
and individual particles have a mass of $8.6\times 10^8\,h^{-1}{\rm
M}_{\odot}$.  This volume is large enough to include interesting rare
objects, but still small enough that the halos of all luminous
galaxies brighter than $0.1 L_\star$ are resolved with at least 100
particles. At the present day, the richest clusters of galaxies
contain about 3 million particles. The gravitational force law is
softened isotropically on a comoving scale of $5\,h^{-1}{\rm kpc}$
(Plummer-equivalent), which may be taken as the spatial resolution
limit of the calculation. Thus, our simulation achieves a dynamic
range of $10^5$ in 3D, and this resolution is available everywhere in
the simulation volume.

Initial conditions were laid down by perturbing a homogeneous,
`glass-like', particle distribution\cite{White1996} with a realization
of a Gaussian random field with the $\Lambda$CDM linear power spectrum
as given by the Boltzmann code {\small CMBFAST}\cite{Seljak1996}. The
displacement field in Fourier space was constructed using the
Zel'dovich approximation, with the amplitude of each random phase mode
drawn from a Rayleigh distribution. The simulation started at redshift
$z=127$ and was evolved to the present using a leapfrog integration
scheme with individual and adaptive timesteps, with up to $11\,000$
timesteps for individual particles.  We stored the full particle data
at 64 output times, each of size 300 GB, giving a raw data volume of
nearly 20 TB.  This allowed the construction of finely resolved
hierarchical merging trees for tens of millions of halos and for the
subhalos that survive within them. A galaxy catalogue for the full
simulation, typically containing $\sim2\times 10^6$ galaxies at $z=0$
together with their full histories, can then be built for any desired
semi-analytic model in a few hours on a high-end workstation.

The semi-analytic model itself can be viewed as a simplified
simulation of the galaxy formation process, where the star formation
and its regulation by feedback processes is parameterised in terms of
simple analytic physical models. These models take the form of
differential equations for the time evolution of the galaxies that
populate each hierarchical merging tree.  In brief, these equations
describe radiative cooling of gas, star formation, growth of
supermassive black holes, feedback processes by supernovae and AGN,
and effects due to a reionising UV background. In addition, the
morphological transformation of galaxies and the process of metal
enrichment are modelled as well.  To make direct contact with
observational data, we apply modern population synthesis models to
predict spectra and magnitudes for the stellar light emitted by
galaxies, also including simplified models for dust obscuration. In
this way we can match the passbands commonly used in observations.

The basic elements of galaxy formation modelling follow previous
studies\cite{WhiteFrenk1991,Kauffmann1993,Cole1994,Baugh1996,Sommerville1999,Kauffmann1999,Springel2001b}
(see also Supplementary Information), but we also use novel approaches
in a number of areas. Of substantial importance is our tracking of
dark matter substructure. This we carry out consistently and with
unprecedented resolution throughout our large cosmological volume,
allowing an accurate determination of the orbits of galaxies within
larger structures, as well as robust estimates of the survival time of
structures infalling into larger objects. Also, we use dark matter
substructure properties, like angular momentum or density profile, to
directly determine sizes of galactic disks and their rotation curves.
Secondly, we employ a novel model for the build-up of a population of
supermassive black holes in the universe. To this end we extend the
quasar model developed in previous work\cite{Kauffmann2000} with a
`radio mode', which describes the feedback activity of central AGN in
groups and clusters of galaxies. While largely unimportant for the
cumulative growth of the total black hole mass density in the
universe, our results show that the radio mode becomes important at
low redshift, where it has a strong impact on cluster cooling
flows. As a result, it reduces the brightness of central cluster
galaxies, an effect that shapes the bright end of the galaxy
luminosity function, bringing our predictions into good agreement with
observation.

\bibliography{main}

\paragraph*{Supplementary Information} accompanies the paper on 
{\bf www.nature.com/nature}.

\vspace*{-0.5cm}\paragraph*{Acknowledgements} We would like to thank the anonymous
referees who helped to improve the paper substantially. The
computations reported here were performed at the {\em Rechenzentrum der
Max-Planck-Gesellschaft} in Garching, Germany.

\vspace*{-0.5cm}\paragraph*{Competing interests} The authors declare that they have no
competing financial interests.

\vspace*{-0.5cm}\paragraph*{Correspondence} and requests for materials should
be addressed to V.S.~(email: vspringel@mpa-garching.mpg.de).


\paragraphfont{\small}
\subsectionfont{\normalsize}

\newcommand\be{\begin{equation}}
\newcommand\ee{\end{equation}}
\renewcommand{\vec}[1]{ {{\bf #1}} } 

\renewcommand{\textfraction}{0}
\renewcommand{\floatpagefraction}{1.0}
\renewcommand{\topfraction}{1.0}
\renewcommand{\bottomfraction}{1.0}

\renewcommand\caption[1]{%
  \myCaption{#1}
}

\title{\vspace*{-1cm}{\Large Simulating the joint evolution of quasars, galaxies\\
    and their large-scale distribution} \vspace*{0.2cm} \\ {\em \large Supplementary Information}\vspace*{0.3cm}}

\author{\parbox{13.5cm}{\small\sffamily%
V.~Springel$^{1}$, %
S.~D.~M.~White$^{1}$, %
A.~Jenkins$^{2}$, %
C.~S.~Frenk$^{2}$, %
N.~Yoshida$^{3}$, %
L.~Gao$^{1}$, %
J.~Navarro$^{4}$, %
R.~Thacker$^{5}$, %
D.~Croton$^{1}$, %
J.~Helly$^{2}$, %
J.~A.~Peacock$^{6}$, %
S.~Cole$^{2}$, %
P.~Thomas$^{7}$, %
H.~Couchman$^{5}$, %
A.~Evrard$^{8}$, %
J.~Colberg$^{9}$ \& %
F.~Pearce$^{10}$}\vspace*{-0.5cm}}

\renewcommand\refname{\large References}

\date{}

\bibliographystyle{naturemag}

\baselineskip14pt 
\setlength{\parskip}{4pt}
\setlength{\parindent}{18pt}%

\twocolumn

\setlength{\footskip}{25pt}
\setlength{\textheight}{670pt}
\setlength{\oddsidemargin}{-8pt}
\setlength{\topmargin}{-41pt}
\setlength{\headsep}{18pt}
\setlength{\textwidth}{469pt}
\setlength{\marginparwidth}{42pt}
\setlength{\marginparpush}{5pt}

\addtolength{\topmargin}{-0.6cm}

\maketitle

\renewcommand{\thefootnote}{\arabic{footnote}}
\footnotetext[1]{\footnotesize Max-Planck-Institute for Astrophysics, Karl-Schwarzschild-Str.~1, 85740 Garching, Germany}
\footnotetext[2]{\footnotesize Institute for Computational Cosmology, Dep. of
    Physics, Univ. of Durham, South Road, Durham  DH1 3LE, UK}
\footnotetext[3]{\footnotesize Department of Physics, Nagoya University, Chikusa-ku, Nagoya 464-8602, Japan}
\footnotetext[4]{\footnotesize Dep. of Physics \& Astron., University of Victoria, Victoria, BC, V8P 5C2, Canada}
\footnotetext[5]{\footnotesize Dep. of Physics \& Astron., McMaster Univ., 1280
  Main St. West, Hamilton, Ontario, L8S 4M1, Canada}
\footnotetext[6]{\footnotesize Institute of Astronomy, University of Edinburgh, Blackford Hill, Edinburgh EH9 3HJ, UK}
\footnotetext[7]{\footnotesize Dep. of Physics \& Astron., University of Sussex, Falmer, Brighton BN1 9QH, UK}
\footnotetext[8]{\footnotesize Dep. of Physics \& Astron., Univ. of Michigan, Ann Arbor, MI 48109-1120, USA}
\footnotetext[9]{\footnotesize Dep. of Physics \& Astron., Univ. of Pittsburgh, 3941 O'Hara Street, Pittsburgh PA 15260, USA}
\footnotetext[10]{\footnotesize Physics and Astronomy Department, Univ. of Nottingham, Nottingham NG7 2RD, UK}

{\bf\small This document provides supplementary information for the
  above article in Nature.  We detail the physical model used to
  compute the galaxy population, and give a short summary of our
  simulation method. Where appropriate, we give further references to
  relevant literature for our methodology. \vspace*{-0.3cm}}

\renewcommand{\thefootnote}{\fnsymbol{footnote}}

\small

\subsection*{Characteristics of the simulation\vspace*{-0.3cm}}

Numerical simulations are a primary theoretical tool to study the
nonlinear gravitational growth of structure in the Universe, and to
link the initial conditions of cold dark matter (CDM) cosmogonies to
observations of galaxies at the present day.  Without direct numerical
simulation, the hierarchical build-up of structure with its
three-dimensional dynamics would be largely inaccessible.

\begin{figure*} 
\begin{center} 
\hspace*{-0.6cm}\resizebox{16cm}{!}{\includegraphics{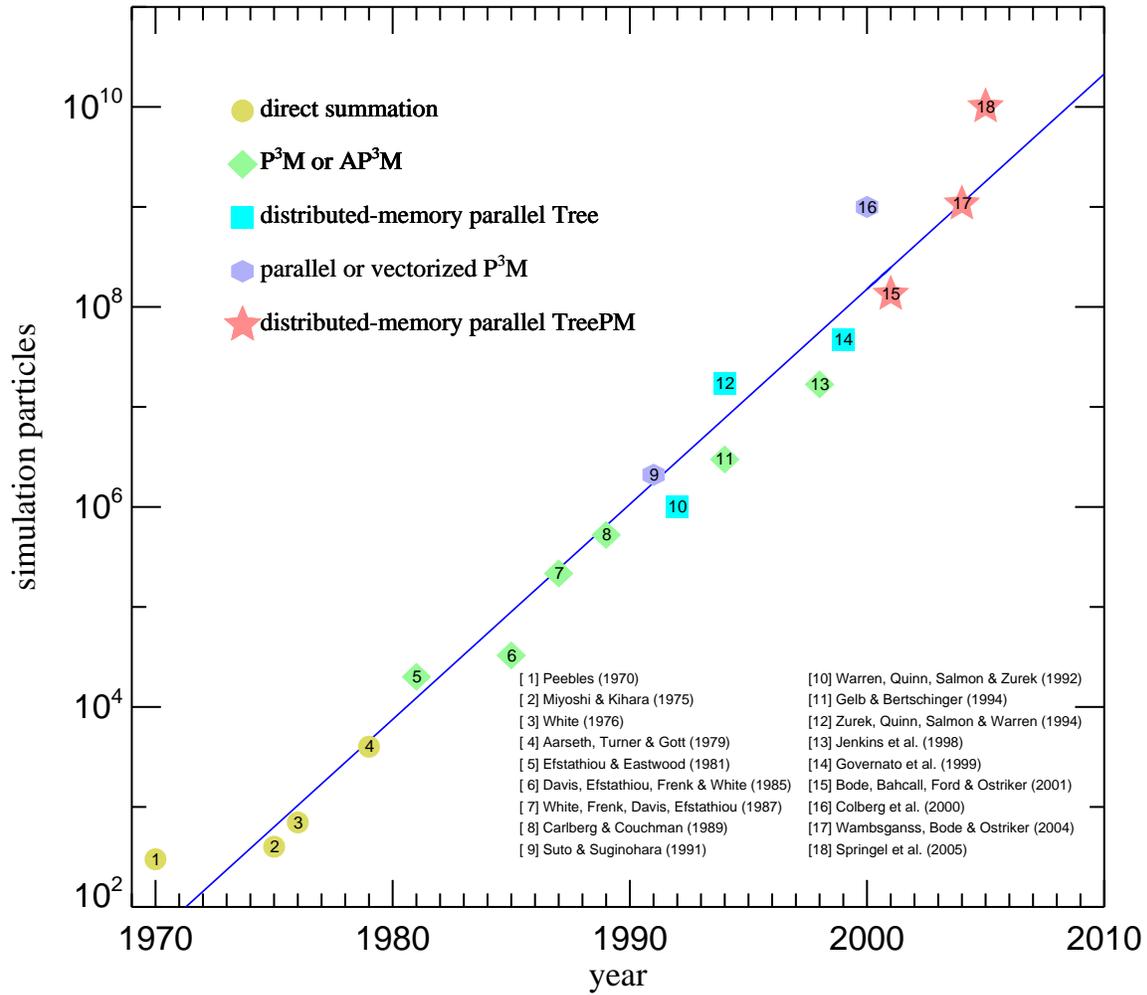}}
\end{center}
\caption{Particle number in high resolution N-body simulations of
  cosmic structure formation as a function of publication
  date\cite{Peebles1970,Miyoshi1975,White1976,Aarseth1979,Efstathiou1981,Davis1985,White1987,Carlberg1989,Suto1991,Warren1992,Gelb1994,Zurek1994,Jenkins1998,Governato1999,Bode2001,Colberg2000,Wambsganss2004}.
  Over the last three decades, the growth in simulation size has been
  exponential, doubling approximately every $\sim 16.5$ months (blue
  line).  Different symbols are used for different classes of
  computational algorithms. The particle-mesh (PM) method combined
  with direct particle-particle (PP) summation on sub-grid scales has
  long provided the primary path towards higher resolution. However,
  due to their large dynamic range and flexibility, tree algorithms
  have recently become competitive with these traditional ${\rm P^3M}$
  schemes, particularly if combined with PM methods to calculate the
  long-range forces. Plain PM
  simulations\cite{Klypin1983,White1983,Centrella1983,Park1990,Bertschinger1991}
  have not been included in this overview because of their much lower
  spatial resolution for a given particle number. Note also that we
  focus on the largest simulations at a given time, so our selection
  of simulations does not represent a complete account of past work on
  cosmological simulations.  \label{FigNvsTime}}
\end{figure*}

Since the dominant mass component, the dark matter, is assumed to
consist of weakly interacting elementary particles that interact only
gravitationally, such simulations use a set of discrete point
particles to represent the collisionless dark matter fluid. This
representation as an N-body system is obviously only a coarse
approximation, and improving its fidelity requires the use of as many
particles as possible while remaining computationally
tractable. Cosmological simulations have therefore always striven to
increase the size (and hence resolution) of N-body computations,
taking advantage of every advance in numerical algorithms and computer
hardware.  As a result, the size of simulations has grown continually
over the last four decades.  Fig.~\ref{FigNvsTime} shows the progress
since 1970.  The number of particles has increased exponentially,
doubling roughly every 16.5 months.  Interestingly, this growth
parallels the empirical `Moore's Law' used to describe the growth of
computer performance in general. Our new simulation discussed in this
paper uses an unprecedentedly large number of $2160^3$ particles, more
than $10^{10}$. We were able to finish this computation in 2004,
significantly ahead of a simple extrapolation of the past growth rate
of simulation sizes. The simulation represented a substantial
computational challenge that required novel approaches both for the
simulation itself, as well as for its analysis. We describe the most
important of these aspects in the following.  As an aside, we note
that extrapolating the remarkable progress since the 1970s for another
three decades, we may expect cosmological simulations with $\sim
10^{20}$ particles some time around 2035.  This would be sufficient to
represent all stars in a region as large as the Millennium volume with
individual particles.

\begin{figure}
\begin{center}
\vspace*{-0.2cm}\resizebox{7.5cm}{!}{\includegraphics{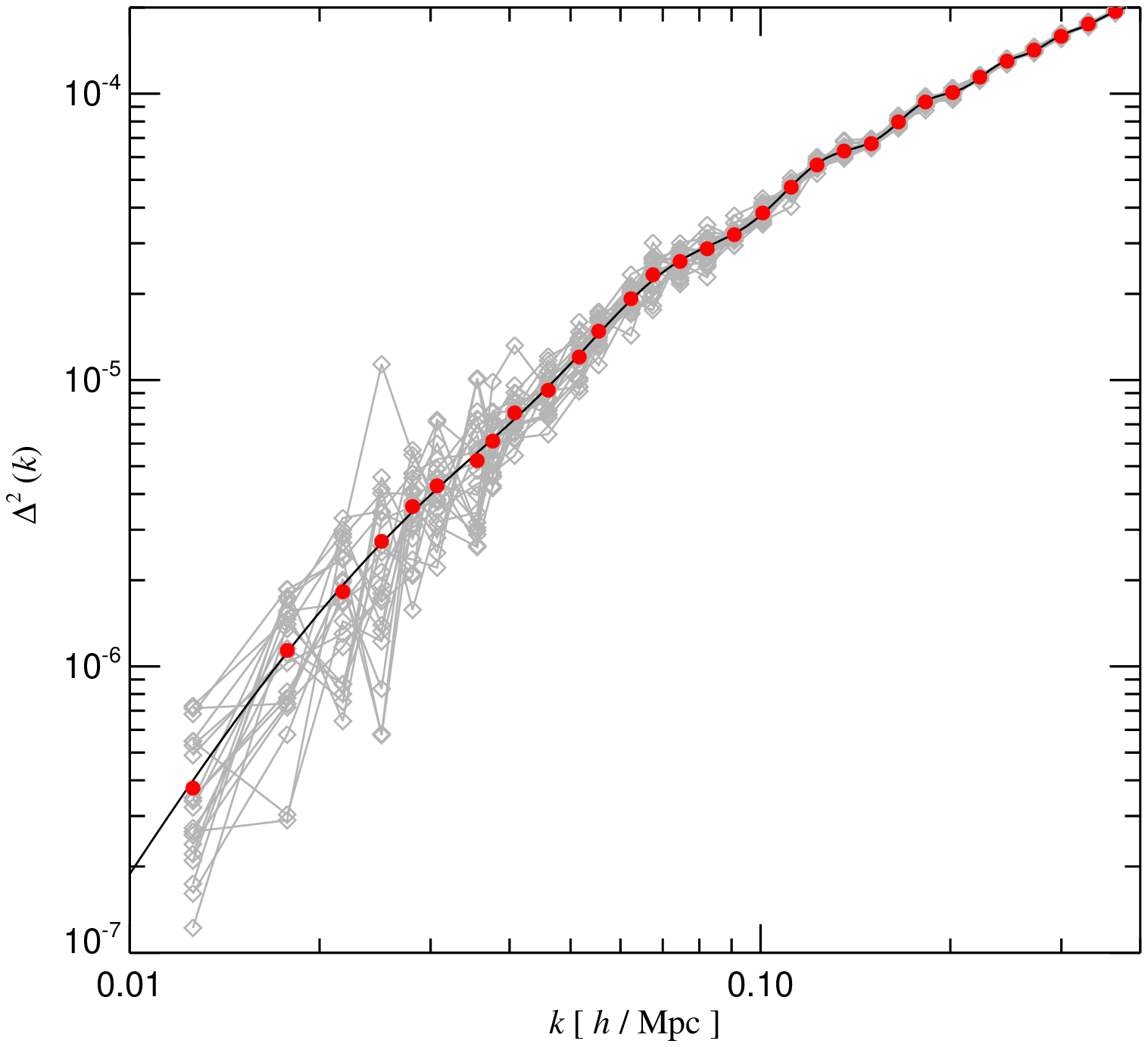}}
\vspace*{-0.2cm}\resizebox{7.5cm}{!}{\includegraphics{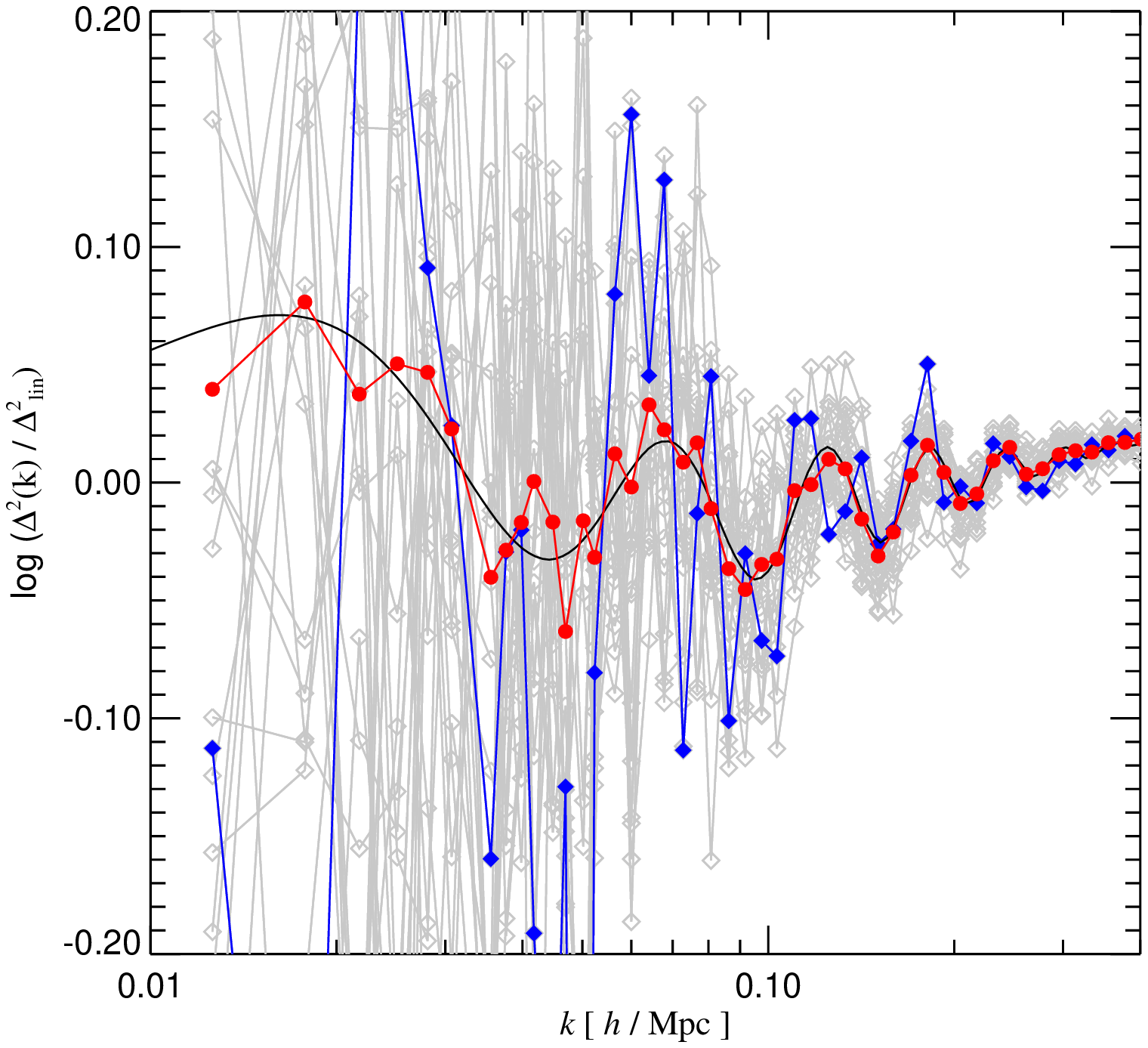}}
\end{center}
\caption{Different realizations of the initial power spectrum. The top
  and bottom panels show measured power-spectra for 20 realizations of
  initial conditions with different random number seeds, together with
  the mean spectrum (red symbols). The latter lies close to the input
  linear power spectrum (black solid line). In the bottom panel, the
  measurements have been divided by a smooth CDM-only power
  spectrum\cite{Bardeen1986} to highlight the acoustic
  oscillations. One of the realizations has been drawn in blue; it
  shows a fluctuation pattern that superficially resembles the pattern
  around the second acoustic peak. However, this is a chance effect;
  the fluctuations of each bin are independent.
\label{FigRealizations}}
\end{figure}

\paragraph*{Initial conditions.}
We used the Boltzmann code {\small CMBFAST}\cite{Seljak1996} to
compute a linear theory power spectrum of a $\Lambda$CDM model with
cosmological parameters consistent with recent constraints from WMAP
and large-scale structure data\cite{Spergel2003,Seljak2004}. We then
constructed a random realization of the model in Fourier space,
sampling modes in a sphere up to the Nyquist frequency of our $2160^3$
particle load. Mode amplitudes $|\delta_{\vec k}|$ were determined by
random sampling from a Rayleigh distribution with second moment equal
to $P(k)=\left<|\delta_{\vec k}|^2\right>$, while phases were chosen
randomly. A high quality random number generator with period $\sim
10^{171}$ was used for this purpose.  We employed a massively parallel
complex-to-real Fourier transform (which requires some care to satisfy
all reality constraints) to directly obtain the resulting displacement
field in each dimension.  The initial displacement at a given particle
coordinate of the unperturbed density field was obtained by tri-linear
interpolation of the resulting displacement field, with the initial
velocity obtained from the Zel'dovich approximation. The latter is
very accurate for our starting redshift of $z=127$. For the initial
unperturbed density field of $2160^3$ particles we used a {\em
glass-like} particle distribution. Such a glass is formed when a
Poisson particle distribution in a periodic box is evolved with the
sign of gravity reversed until residual forces have dropped to
negligible levels\cite{White1996}.  For reasons of efficiency, we
replicated a $270^3$ glass file 8 times in each dimension to generate
the initial particle load.  The Fast Fourier Transforms (FFT) required
to compute the displacement fields were carried out on a $2560^3$ mesh
using 512 processors and a distributed-memory code. We deconvolved the
input power spectrum for smoothing effects due to the interpolation
off this grid.

We note that the initial random number seed was picked in an
unconstrained fashion. Due to the finite number of modes on large
scales and the Rayleigh-distribution of mode amplitudes, the mean
power of the actual realization in each bin is expected to scatter
around the linear input power spectrum. Also, while the expectation
value $\left<|\delta_{\vec k}|^2\right>$ is equal to the input power
spectrum, the median power per mode is biased low due to the
skew-negative distribution of the mode amplitudes. Hence, in a given
realization there are typically more points lying below the input
power spectrum than above it, an effect that quickly becomes
negligible as the number of independent modes in each bin becomes
large.  We illustrate this in the top panel of
Figure~\ref{FigRealizations}, where 20 realizations for different
random number seeds of the power spectrum on large scales are shown,
together with the average power in each bin. Our particular
realization for the Millennium Simulation corresponds to a slightly
unlucky choice of random number seed in the sense that the
fluctuations around the mean input power in the region of the second
peak seem to resemble the pattern of the acoustic oscillations (see
the top left panel of Figure~6 in our Nature article).  However, we
stress that the fluctuations in these bins are random and
uncorrelated, and that this impression is only a chance effect. In the
bottom panel of Figure~\ref{FigRealizations}, we redraw the measured
power spectra for the 20 random realizations, this time normalised to
a smooth CDM power spectrum without acoustic oscillations in order to
highlight the baryonic `wiggles'. We have drawn one of the 20
realizations in blue. It is one that resembles the pattern of
fluctuations seen in the Millennium realization quite closely while
others scatter quite differently, showing that such deviations are
consistent with the expected statistical distribution.

\paragraph*{Dynamical evolution.}

The evolution of the simulation particles under gravity in an
expanding background is governed by the Hamiltonian \be H= \sum_i
\frac{\vec{p}_i^2}{2\,m_i\, a(t)^2} + \frac{1}{2}\sum_{ij}\frac{m_i
m_j \,\varphi(\vec{x}_i-\vec{x}_j)}{a(t)}, \label{eqHamil} \ee where
$H=H(\vec{p}_1,\ldots,\vec{p}_N,\vec{x}_1,\ldots,\vec{x}_N, t)$.  The
$\vec{x}_i$ are comoving coordinate vectors, and the corresponding
canonical momenta are given by $\vec{p}_i=a^2 m_i \dot\vec{x}_i$.  The
explicit time dependence of the Hamiltonian arises from the evolution
$a(t)$ of the scale factor, which is given by the Friedman-Lemaitre
model that describes the background cosmology. Due to our assumption
of periodic boundary conditions for a cube of size $L^3$, the
interaction potential $\varphi(\vec{x})$ is the solution of \be
\nabla^2 \varphi(\vec{x}) = 4\pi G \left[ - \frac{1}{L^3} +
\sum_{\vec{n}} \delta_\epsilon(\vec{x}-\vec{n}L)\right], \label{eqpot}
\ee where the sum over $\vec{n}=(n_1, n_2, n_3)$ extends over all
integer triplets.  The density distribution function
$\delta_\epsilon(\vec{x})$ of a single particle is spread over a
finite scale $\epsilon$, the gravitational softening length.  The
softening is necessary to make it impossible for hard binaries to form
and to allow the integration of close particle encounters with
low-order integrators.  We use a spline kernel to soften the point
mass, given by $\delta_\epsilon(\vec{x}) = W(|\vec{x}|/2.8\epsilon)$,
where $W(r) = 8 (1- 6 r^2 + 6 r^3)/\pi$ for $0\le r<1/2$, $W(r) = {16}
(1- r)^3/\pi$ for $1/2 \le r<1$, and $W(r)=0$ otherwise.  For this
choice, the Newtonian potential of a point mass at zero lag in
non-periodic space is $-G\,m/\epsilon$, the same as for a
`Plummer-sphere' of size $\epsilon$, and the force becomes fully
Newtonian for separations larger than $2.8\epsilon$. We took
$\epsilon=5\,h^{-1}{\rm kpc}$, about $46.3$ times smaller than the
mean particle separation. Note that the mean density is subtracted in
equation (\ref{eqpot}), so the solution of the Poisson equation
corresponds to the {\em peculiar potential}, where the dynamics of the
system is governed by $\nabla^2 \phi(\vec{x}) = 4\pi G
[\rho(\vec{x})-\overline\rho]$.

The equations of motion corresponding to equation (\ref{eqHamil}) are
$\sim 10^{10}$ simple differential equations, which are however
coupled tightly by the mutual gravitational forces between the
particles. An accurate evaluation of these forces (the `right hand
side' of the equations) is computationally very expensive, even when
force errors up to $\sim 1\%$ can be tolerated, which is usually the
case in collisionless dynamics\cite{Hernquist1993}.  We have written a
completely new version of the cosmological simulation code {\small
GADGET}\cite{Springel2001} for this purpose.  Our principal
computational technique for the gravitational force calculation is a
variant of the `TreePM' method\cite{Xu1995,Bode2000,Bagla2002}, which
uses a hierarchical multipole expansion\cite{Barnes1986} (a `tree'
algorithm) to compute short-range gravitational forces and combines
this with a more traditional particle-mesh (PM)
method\cite{Hockney1981} to determine long-range gravitational forces.
This combination allows for a very large dynamic range and high
computational speed even in situations where the clustering becomes
strong. We use an explicit force-split\cite{Bagla2002} in
Fourier-space, which produces a highly isotropic force law and
negligible force errors at the force matching scale. The algorithms in
our code are specially designed for massively parallel operation and
contain explicit communication instructions such that the code can
work on computers with distributed physical memory, a prerequisite for
a simulation of the size and computational cost of the Millennium Run.

\begin{figure*}
\begin{center}
\resizebox{15cm}{!}{\includegraphics{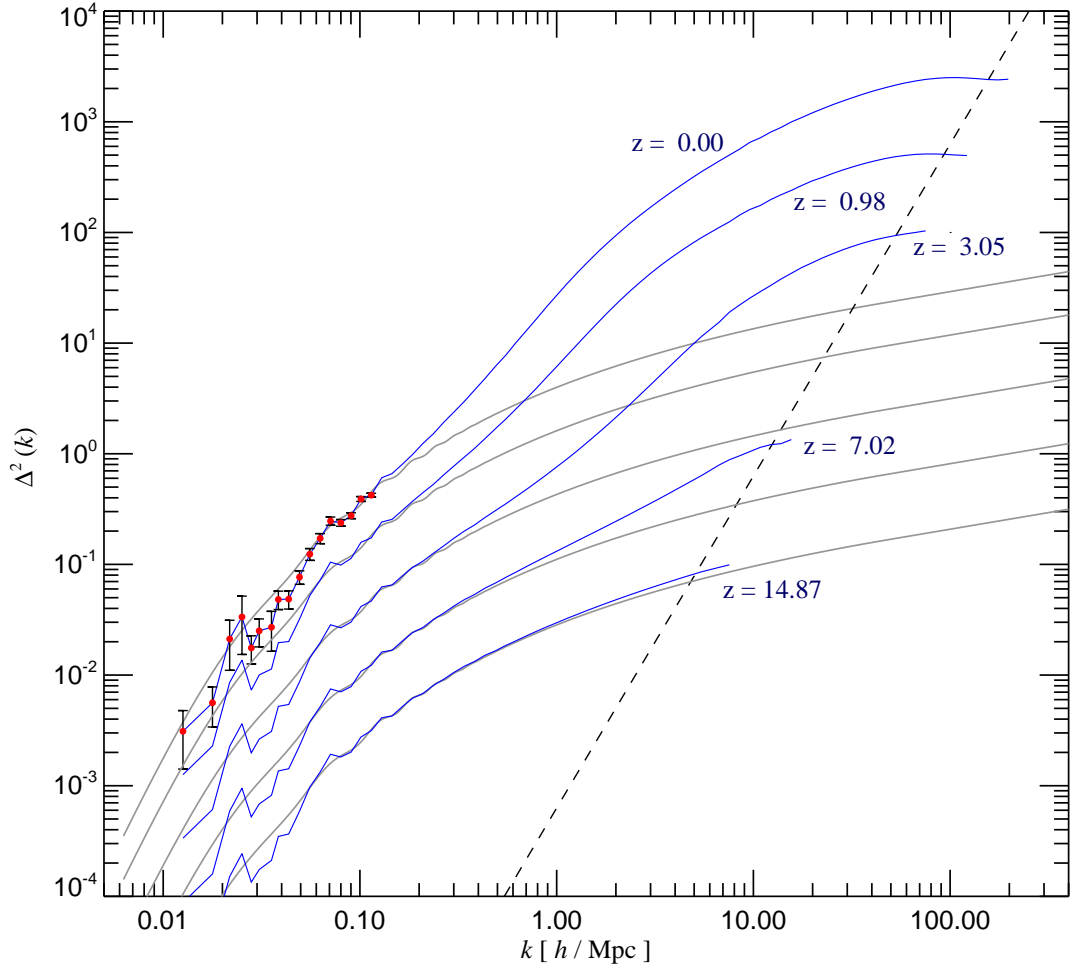}}
\end{center}
\caption{The power spectrum of the dark matter distribution in the
Millennium Simulation at various epochs (blue lines). The gray lines
show the power spectrum predicted for linear growth, while the dashed
line denotes the shot-noise limit expected if the simulation particles
are a Poisson sampling from a smooth underlying density field. In
practice, the sampling is significantly sub-Poisson at early times and
in low density regions, but approaches the Poisson limit in nonlinear
structures. Shot-noise subtraction allows us to probe the spectrum
slightly beyond the Poisson limit. Fluctuations around the linear
input spectrum on the largest scales are due to the small number of
modes sampled at these wavelengths and the Rayleigh distribution of
individual mode amplitudes assumed in setting up the initial
conditions.  To indicate the bin sizes and expected sample variance on
these large scales, we have included symbols and error bars in the
$z=0$ estimates. On smaller scales, the statistical error bars are
negligibly small.
\label{FigPowerSpec}}
\end{figure*}

For the tree-algorithm, we first decompose the simulation volume
spatially into compact {\em domains}, each served by one
processor. This domain decomposition is done by dividing a space
filling Peano-Hilbert curve into segments. This fractal curve visits
each cell of a fiducial grid of $1024^3$ cells overlayed over the
simulation exactly once. The decomposition tries to achieve a
work-load balance for each processor, and evolves over time as
clustering progresses. Using the Peano-Hilbert curve guarantees that
domain boundaries are always parallel to natural tree-node boundaries,
and thanks to its fractal nature provides for a small
surface-to-volume ratio for all domains, such that communication with
neighbouring processors during the short-range tree force computation
can be minimised. Our tree is fully threaded (i.e.~its leaves are
single particles), and implements an oct-tree structure with monopole
moments only. The cell-opening criterion was
relative\cite{Salmon1994}; a multipole approximation was accepted if
its conservatively estimated error was below $0.5\%$ of the total
force from the last timestep. In addition, nodes were always opened
when the particle under consideration lay inside a 10\% enlarged outer
node boundary. This procedure gives forces with typical errors well
below $0.1\%$.

For the PM algorithm, we use a parallel Fast Fourier Transform
(FFT)\footnote[1]{Based on the www.fftw.org libraries of MIT.} to
solve Poisson's equation.  We used a FFT mesh with $2560^3$ cells,
distributed into 512 slabs of dimension $5\times 2560\times 2560$ for
the parallel transforms.  After clouds-in-cells (CIC) mass assignment
to construct a density field, we invoke a real-to-complex transform to
convert to Fourier space. We then multiplied by the Greens function of
the Poisson equation, deconvolved for the effects of the CIC and the
trilinear interpolation that is needed later, and applied the
short-range filtering factor used in our TreePM formulation (the short
range forces suppressed here are exactly those supplied by the
tree-algorithm). Upon transforming back we obtained the gravitational
potential. We then applied a four-point finite differencing formula to
compute the gravitational force field for each of the three coordinate
directions.  Finally, the forces at each particle's coordinate were
obtained by trilinear interpolation from these fields.

A particular challenge arises due to the different data layouts needed
for the PM and tree algorithms. In order to keep the required
communication and memory overhead low, we do not swap the particle
data between the domain and slab decompositions.  Instead, the
particles stay in the domain decomposition needed by the tree, and
each processor constructs patches of the density field for all the
slabs on other processors which overlap its local domain.  In this
way, each processor communicates only with a small number of other
processors to establish the binned density field on the slabs.
Likewise, the slab-decomposed potential field is transfered back to
processors so that a local region is formed covering the local domain,
in addition to a few ghost cells around it, such that the finite
differencing of the potential can be carried out for all interior
points.

Timestepping was achieved with a symplectic leap-frog scheme based on
a split of the potential energy into a short-range and long-range
component. The short-range dynamics was then integrated by subcycling
the long-range step\cite{Duncan1998}. Hence, while the short-range
force had to be computed frequently, the long-range FFT force was
needed only comparatively infrequently. More than 11000 timesteps in
total were carried out for the simulation, using individual and
adaptive timesteps\footnote[2]{Allowing adaptive changes of timesteps
formally breaks the symplectic nature of our integration scheme, which
is however not a problem for the dynamics we follow here.} for the
particles. A timestep of a particle was restricted to be smaller than
$\Delta t = \sqrt{2 \eta \epsilon/|\vec{a}|}$, where $\vec{a}$ is a
particle's acceleration and $\eta=0.02$ controls the integration
accuracy. We used a binary hierarchy of timesteps to generate a
grouping of particles onto timebins.

The memory requirement of the code had to be aggressively optimised in
order to make the simulation possible on the IBM p690 supercomputer
available to us. The total aggregated memory on the 512 processors was
1 TB, of which about 950~GB could be used freely by an application
program. In our code {\small {\em Lean}-GADGET-2} produced for the
Millennium Simulation, we needed about 400~GB for particle storage and
300~GB for the fully threaded tree in the final clustered particle
state, while the PM algorithm consumed in total about 450~GB in the
final state (due to growing variations in the volume of domains as a
result of our work-load balancing strategy, the PM memory requirements
increase somewhat with time). Note that the memory for tree and PM
computations is not needed concurrently, and this made the simulation
feasible.  The peak memory consumption per processor reached 1850~MB
at the end of our simulation, rather close to the maximum possible of
1900~MB.

\begin{figure}
\begin{center}
\resizebox{8.5cm}{!}{\includegraphics{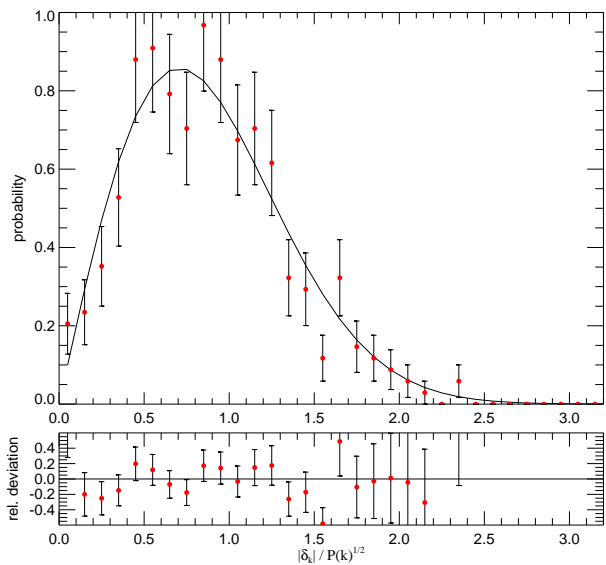}}\vspace*{-0.3cm}
\end{center}
\caption{Measured distribution of mode amplitudes in the Millennium
  Simulation at redshift $z=4.9$. Only modes in the $k$-range
    $0.03\,h/{\rm Mpc} < k < 0.07\,h/{\rm Mpc}$ are included (in total
    341 modes), with their amplitude normalised to the square root of
    the expected linear power spectrum at that redshift. The
    distribution of modes follows the expected Rayleigh distribution
    very well. The bottom panel shows the relative deviations of the
    measurements from this distribution, which are in line with the
    expected statistical scatter.
\label{FigRayleigh}}
\end{figure}

\paragraph*{On the fly analysis.}

With a simulation of the size of the Millennium Run, any non-trivial
analysis step is demanding. For example, measuring the dark matter
mass power spectrum over the full dynamic range of the simulation
volume would require a 3D FFT with $\sim 10^5$ cells per dimension,
which is unfeasible at present.  In order to circumvent this problem,
we employed a two stage procedure for measuring the power spectrum
where a ``large-scale'' and a ``small-scale'' measurement were
combined. The former was computed with a Fourier transform of the
whole simulation box, while the latter was constructed by folding the
density field back onto itself\cite{Jenkins1998}, assuming periodicity
for a fraction of the box.  The self-folding procedure leads to a
sparser sampling of Fourier space on small scales, but since the
number of modes there is large, an accurate small-scale measurement is
still achieved.  Since the PM-step of the simulation code already
computes an FFT of the whole density field, we took advantage of this
and embedded a measurement of the power spectrum directly into the
code. The self-folded spectrum was computed for a 32 times smaller
periodic box-size, also using a $2560^3$ mesh, so that the power
spectrum measurement effectively corresponded to a $81920^3$ mesh. We
have carried out a measurement each time a simulation snapshot was
generated and saved on disk. In Figure~\ref{FigPowerSpec}, we show the
resulting time evolution of the {\it dark matter} power spectrum in
the Millennium Simulation.  On large scales and at early times, the
mode amplitudes grow linearly, roughly in proportion to the
cosmological expansion factor. Nonlinear evolution accelerates the
growth on small scales when the dimensionless power $\Delta^2(k)= k^3
P(k)/(2\pi^2)$ approaches unity; this regime can only be studied
accurately using numerical simulations. In the Millennium Simulation,
we are able to determine the nonlinear power spectrum over a larger
range of scales than was possible in earlier work\cite{Jenkins1998},
almost five orders of magnitude in wavenumber $k$.

On the largest scales, the periodic simulation volume encompasses only
a relatively small number of modes and, as a result of the Rayleigh
amplitude sampling that we used, these (linear) scales show
substantial random fluctuations around the mean expected power.  This
also explains why the mean power in the $k$-range $0.03\,h/{\rm Mpc} <
k < 0.07\,h/{\rm Mpc}$ lies below the linear input power.  In
Figure~\ref{FigRayleigh}, we show the actual distribution of
normalised mode amplitudes, $\sqrt{|\delta_\vec{k}|^2 / P(k)}$,
measured directly for this range of wavevectors in the Millennium
Simulation at redshift $z=4.9$. We see that the distribution of mode
amplitudes is perfectly consistent with the expected underlying
Rayleigh distribution.

Useful complementary information about the clustering of matter in
real space is provided by the two-point correlation function of dark
matter particles.  Measuring it involves, in principle, simply
counting the number of particle pairs found in spherical shells around
a random subset of all particles.  Naive approaches to determine these
counts involve an $N^2$-scaling of the operation count and are
prohibitive for our large simulation. We have therefore implemented
novel parallel methods to measure the two-point function accurately,
which we again embedded directly into the simulation code, generating
a measurement automatically at every output. Our primary approach to
speeding up the pair-count lies in using the hierarchical grouping
provided by the tree to search for particles around a randomly
selected particle. Since we use logarithmic radial bins for the pair
counts, the volume corresponding to bins at large radii is
substantial. We use the tree for finding neighbours with a
range-searching technique.  In carrying out the tree-walk, we check
whether a node falls fully within the volume corresponding to a
bin. In this case, we terminate the walk along this branch of the tree
and simply count all the particles represented by the node at once,
leading to a significant speed-up of the measurement.

Finally, the exceptionally large size of the simulation prompted us to
develop new methods for computing friends-of-friends (FOF) group
catalogues in parallel and on the fly.  The FOF groups are defined as
equivalence classes in which any pair of particles belongs to the same
group if their separation is less than 0.2 of the mean particle
separation. This criterion combines particles into groups with a mean
overdensity that corresponds approximately to the expected density of
virialised groups. Operationally, one can construct the groups by
starting from a situation in which each particle is first in its own
single group, and then testing all possible particle pairs; if a close
enough pair is found whose particles lie in different groups already
present, the groups are linked into a common group.  Our algorithm
represents groups as link-lists, with auxiliary pointers to a list's
head, tail, and length. In this way we can make sure that, when groups
are joined, the smaller of two groups is always attached to the tail
of the larger one. Since each element of the attached group must be
visited only once, this procedure avoids a quadratic contribution to
the operation count proportional to the group size when large groups
are built up.  Our parallel algorithm works by first determining the
FOF groups on local domains, again exploiting the tree for range
searching techniques, allowing us to find neighbouring particles
quickly.  Once this first step of group finding for each domain is
finished, we merge groups that are split by the domain decomposition
across two or several processors. As groups may in principle percolate
across several processors, special care is required in this step as
well. Finally, we save a group catalogue to disk at each output,
keeping only groups with at least 20 particles.

In summary, the simulation code evolved the particle set for more than
11000 timesteps, producing 64 output time slices each of about 300 GB.
Using parallel I/O techniques, each snapshot could be written to disk
in about 300 seconds.  Along with each particle snapshot, the
simulation code produced a FOF group catalogue, a power spectrum
measurement, and a two-point correlation function measurement.
Together, over $\sim20$~TB of data were generated by the
simulation. The raw particle data of each output was stored in a
special way (making use of a space-filling curve), which allows rapid
direct access to subvolumes of the particle data. The granularity of
these subvolumes corresponds to a fiducial $256^3$ mesh overlayed over
the simulation volume, such that the data can be accessed randomly in
pieces of $\sim 600$ particles on average.  This storage scheme is
important to allow efficient post-processing, which cannot make use of
an equally powerful supercomputer as the simulation itself.

\begin{figure*}
\begin{center}\resizebox{15cm}{!}{\includegraphics{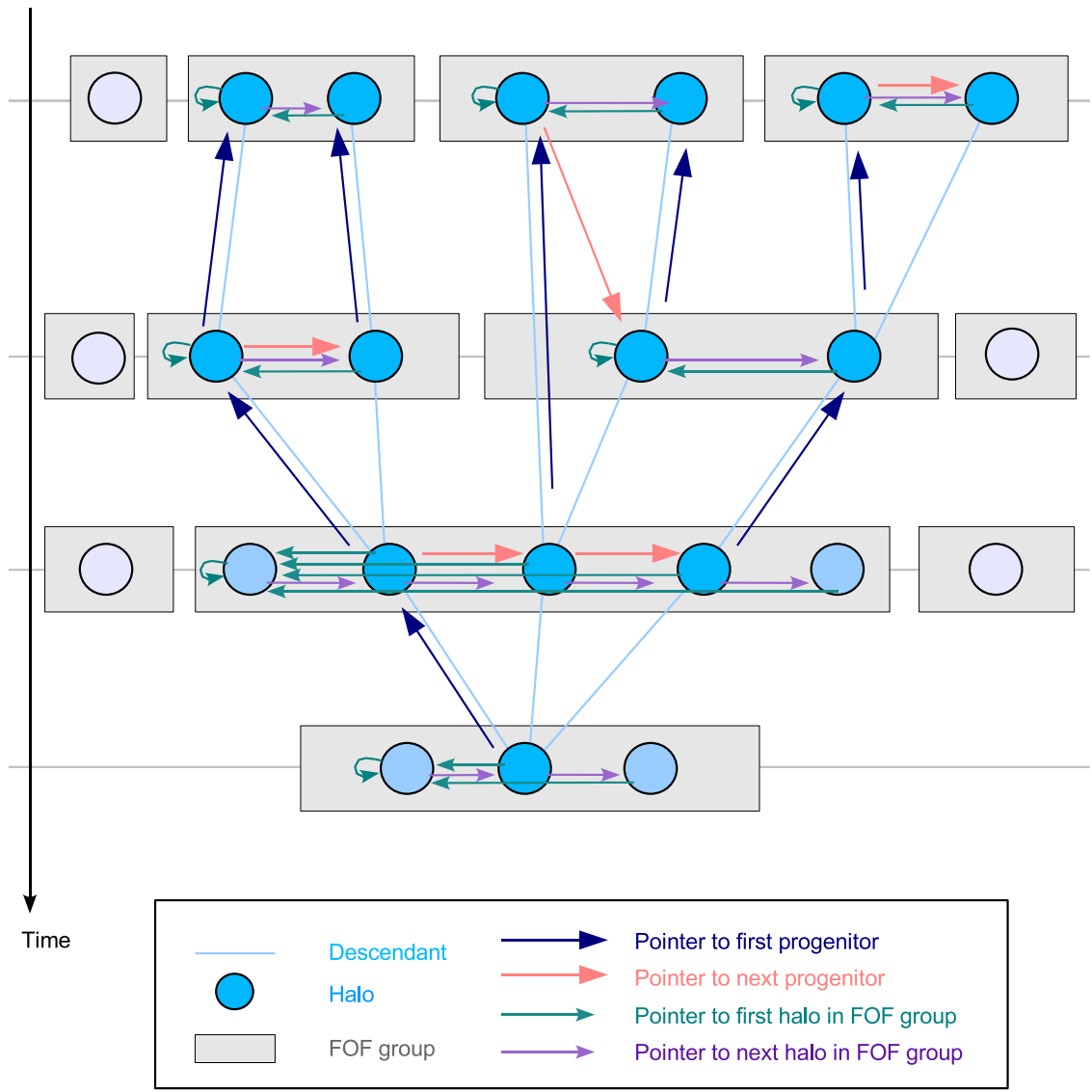}}
\end{center}
\caption{Schematic organisation of the merger tree in
  the Millennium Run. At each output time, FOF groups are identified
  which contain one or several (sub)halos. The merger tree connects
  these halos. The FOF groups play no direct role, except that the
  largest halo in a given FOF group is the one which may develop a
  cooling flow according to the physical model for galaxy formation
  implemented for the trees. To facilitate the latter, a number of
  pointers for each halo are defined. Each halo knows its descendant,
  and its most massive progenitor. Possible further progenitors can be
  retrieved by following the chain of `next progenitors'. In a similar
  fashion, all halos in a given FOF group are linked together.
\label{FigMergTree}}
\end{figure*}

\subsection*{Postprocessing of the simulation data\vspace*{-0.3cm}}

\paragraph*{Substructure analysis.}

High-resolution simulations like the present one exhibit a rich
substructure of gravitationally bound dark matter subhalos orbiting
within larger virialised structures\cite{Ghigna1998}. The FOF group
finder built into the simulation code is able to identify the latter,
but not the `subhalos'. In order to follow the fate of infalling halos
and galaxies more reliably, we therefore determine dark matter
substructures for all identified FOF halos.  We accomplish this with
an improved and extended version of the {\small SUBFIND}
algorithm\cite{Springel2001b}. This computes an adaptively smoothed
dark matter density field using a kernel-interpolation technique, and
then exploits the topological connectivity of excursion sets above a
density threshold to identify substructure candidates. Each
substructure candidate is subjected to a gravitational unbinding
procedure. If the remaining bound part has more than 20 particles, the
subhalo is kept for further analysis and some basic physical
properties (angular momentum, maximum of its rotation curve, velocity
dispersion, etc.) are determined. An identified subhalo was extracted
from the FOF halo, so that the remainder formed a featureless
`background' halo which was also subjected to an unbinding
procedure. The required computation of the gravitational potential for
the unbinding was carried out with a tree algorithm similar to the one
used in the simulation code itself.

Finally, we also compute a virial mass estimate for each FOF halo in
this analysis step, using the spherical-overdensity approach and the
minimum of the gravitational potential within the group as the central
point.  We identified $17.7\times 10^6$ FOF groups at $z=0$, down from
a maximum of $19.8\times 10^6$ at $z=1.4$, where the groups are more
abundant yet smaller on average.  At $z=0$, we found a total of
$18.2\times 10^6$ subhalos, and the largest FOF group contained 2328
of them.

\paragraph*{Merger tree definition and construction.}

Having determined all halos and subhalos at all output times, we
tracked these structures over time, i.e.~we determined the
hierarchical merging trees that describe in detail how structures
build up over cosmic time. These trees are the key information needed
to compute physical models for the properties of the associated galaxy
population.

Because structures merge hierarchically in CDM universes, a given halo
can have several progenitors but, in general, it has only one
descendant because the cores of virialised dark matter structures do
not split up into two or more objects. We therefore based our merger
tree construction on the determination of a unique descendant for any
given halo. This is, in fact, already sufficient to define the merger
tree construction, since the progenitor information then follows
implicitly.

To determine the appropriate descendant, we use the unique IDs that
label each particle and track them between outputs. For a given halo,
we find all halos in the subsequent output that contain some of its
particles.  We count these particles in a weighted fashion, giving
higher weight to particles that are more tightly bound in the halo
under consideration. In this way, we give preference to tracking the
fate of the inner parts of a structure, which may survive for a long
time upon infall into a bigger halo, even though much of the mass in
the outer parts can be quickly stripped. The weighting is facilitated
by the fact that the results of the {\small SUBFIND} analysis are
stored in order of increasing total binding energy, i.e.~the most
bound particle of a halo is always stored first.  Once these weighted
counts are determined for each potential descendant, we select the one
with the highest count as the descendant. As an additional refinement
(which is not important for any of our results), we have allowed some
small halos to skip one snapshot in finding a descendant. This deals
with cases where we would otherwise lose track of a structure that
temporarily fluctuates below our detection threshold.

In Figure~\ref{FigMergTree}, we show a schematic representation of the
merger tree constructed in this way. The FOF groups are represented at
different times with boxes each of which contains one or more
(sub)halos. For each halo, a unique descendant is known, and there are
link-list structures that allow the retrieval of all progenitors of a
halo, or of all other halos in the same FOF group. Not all the trees
in the simulation volume are connected with each other. Instead, there
are $14.4\times 10^6$ separate trees, each essentially describing the
formation history of the galaxies contained in a FOF halo at the
present time. The correspondence between trees and FOF halos is not
exactly one-to-one because some small FOF halos did not contain a
bound subhalo and were dropped, or because some FOF halos can be
occasionally linked by feeble temporary particle bridges which then
also combines their corresponding trees. We have stored the resulting
tree data on a per-tree basis, so that the physical model for galaxy
formation can be computed sequentially for all the trees individually,
instead of having to apply the model in one single step. The latter
would have been impossible, given that the trees contain a total of
around 800 million halos.

\subsection*{Physical model for galaxy formation\vspace*{-0.3cm}}

`Semi-analytic' models of galaxy formation were first proposed more
than a decade ago\cite{White1991}. They have proven to be a very
powerful tool for advancing the theory of galaxy
formation\cite{Kauffmann1993,Cole1994,Kauffmann1996,Kauffmann1997,Baugh1998,Sommerville1999,Cole2000,Benson2002},
even though much of the detailed physics of star formation and its
regulation by feedback processes has remained poorly understood. The
term `semi-analytic' conveys the notion that while in this approach
the physics is parameterised in terms of simple analytic models,
following the dark matter merger trees over time can only be carried
out numerically. Semi-analytic models are hence best viewed as
simplified simulations of the galaxy formation process.  While the
early work employed Monte-Carlo realizations of dark matter
trees\cite{Kauffmann1993tree,Somerville1999tree}, more recent work is
able to measure the merging trees directly from numerical dark matter
simulations\cite{Kauffmann1999}. In the most sophisticated version of
this technique, the approach is extended to include dark matter
substructure information as well\cite{Springel2001b}. This offers
substantially improved tracking of the orbits of infalling
substructure and of their lifetimes. In the Millennium Simulation, we
have advanced this method further still, using improved substructure
finding and tracking methods, allowing us fully to exploit the
superior statistics and information content offered by the underlying
high-resolution simulation.

Our semi-analytic model integrates a number of differential equations
for the time evolution of the galaxies that populate each hierarchical
merging tree.  In brief, these equations describe radiative cooling of
gas, star formation, the growth of supermassive black holes, feedback
processes by supernovae and AGN, and effects due to a reionising UV
background. Morphological transformation of galaxies and processes of
metal enrichment are modelled as well.  Full details of the scheme
used to produce specific models shown in our Nature article will be
provided in a forthcoming publication\cite{Croton2005}, but we here
include a very brief summary of the most important aspects of the
model. Note, however, that this is just one model among many that can
be implemented in post-processing on our stored Millennium Run
data-structures. A prime goal of our project is to evaluate such
schemes against each other and against the observational data in order
to understand which processes determine the various observational
properties of the galaxy population.

\paragraph*{Radiative cooling and star formation.}

We assume that each virialised dark matter halo contains (initially) a
baryonic fraction equal to the universal fraction of baryons,
$f_b=0.17$, which is consistent with WMAP and Big-Bang nucleosynthesis
constraints. A halo may lose some gas temporarily due to heating by a
UV background or other feedback processes, but this gas is assumed to
be reaccreted once the halo has grown in mass sufficiently. The
influence of the UV background is directly taken into account as a
reduction of the baryon fraction for small halos, following fitting
functions obtained from detailed hydrodynamical
models\cite{Gnedin2000,Kravtsov2004}.

We distinguish between cold condensed gas in the centre of halos
(forming the interstellar medium), and hot gas in the diffuse
atmospheres of halos. The latter has a temperature equal to the virial
temperature of the halo, and emits bremsstrahlung and line radiation.
The corresponding cooling rate is estimated following standard
parameterisations\cite{White1991,Springel2001b}, which have been shown
to provide accurate matches to direct hydrodynamical simulations of
halo formation including radiative
cooling\cite{Yoshida2002,Helly2003}. We note that, following the
procedures established already in reference\cite{White1991}, the
cooling model accounts for a distinction between a cold infall regime
and cooling out of a hot atmosphere. The transition is at a mass scale
close to that found in detailed analytic calculations of the cooling
process\cite{Forcado1997,Birnboim2003} and in recent hydrodynamical
simulations\cite{Keres2004}.

The cooling gas is assumed to settle into a disk supported by
rotation. We directly estimate the disk size based on the spin
parameter of the hosting dark matter halo\cite{Mo1998}. Once the gas
surface density exceeds a critical threshold motivated by
observations\cite{Kauffmann1996b}, we assume that star formation
proceeds in the disk, with an efficiency of order $\simeq 10\%$ on a disk
dynamical time. This parameterisation reproduces the phenomenological
laws of star formation in observed disk
galaxies\cite{Kennicutt1989,Kennicutt1998} and the observed gas
fractions at low redshift.

Supernova explosions associated with short-lived massive stars are
believed to regulate star formation in galaxies, particularly in small
systems with shallow potential wells\cite{Dekel1986}. Observations
suggest that supernovae blow gas out of star-forming disks, with a
rate that is roughly proportional to the total amount of stars
formed\cite{Martin1999}. We adopt this observational scaling, and
estimate how much of this gas can join the hot halo of the galaxy
given the total amount of energy released by the supernovae, and how
much may be blown out of the halo entirely. The efficiency of such
mass-loss is a strong function of the potential well depth of the
galaxy. In our model, small galaxies may blow away their remaining gas
entirely in an intense burst of star formation, while large galaxies
do not exhibit any outflows.

\paragraph*{Morphological evolution.}

We characterise galaxy morphology by a simple bulge-to-disk ratio
which can be transformed into an approximate Hubble type according
observational trends\cite{Simien1986}. While the generic mode of gas
cooling leads to disk formation, we consider two possible channels for
the formation of bulges: secular evolution due to disk instabilities,
or as a consequence of galaxy merger events.

Secular evolution can trigger bar and bulge formation in disk
galaxies. We invoke simple stability arguments for self-gravitating
stellar disks\cite{Mo1998} to determine the mass of stars that needs
to be put into a nuclear bulge component to render the stellar disk
stable.

Galaxy mergers are described by the halo merger tree constructed from
the simulation, augmented with a timescale for the final stages of a
merger whenever we lose track of a substructure due to finite spatial
and time resolution. We then estimate the remaining survival time in a
standard fashion based on the dynamical friction timescale. We use the
mass ratio of two merging galaxies to distinguish between two classes
of mergers. {\em Minor mergers} involve galaxies with mass ratio less
than $0.3$. In this case, we assume that the disk of the larger galaxy
survives, while the merging satellite becomes part of the bulge
component. For larger mass ratios, we assume a {\em major merger}
takes place, leading to destruction of both disks, and reassembly of
all stars in a common spheroid. Such an event is the channel through
which pure elliptical galaxies can form. The cold gas in the satellite
of a minor merger, or the cold gas in both galaxies of a major merger,
is assumed to be partially or fully consumed in a nuclear starburst in
which additional bulge stars are formed.  The detailed
parameterisation of such induced starbursts follows results obtained
from systematic parameter studies of hydrodynamical galaxy collision
simulations\cite{Mihos1994,Mihos1996,Cox2004}.

\paragraph*{Spectrophotometric modelling.}

To make direct contact with observational data, it is essential to
compute spectra and magnitudes for the model galaxies, in the
passbands commonly used in observations. Modern population synthesis
models allow an accurate prediction of spectrophotometric properties
of stellar populations as a function of age and
metallicity\cite{Bruzual1993,Bruzual2003}. We apply such a
model\cite{Bruzual2003} and compute magnitudes in a number of
passbands separately for both bulge and disk components and in both
rest- and observer-frames. Dust obscuration effects are difficult to
model in general and present a major source of uncertainty, especially
for simulated galaxies at high redshift. We apply a rather simple
plane-parallel slab dust model\cite{Kauffmann1999}, as a first-order
approximation to dust extinction.

Metal enrichment of the diffuse gas component can also be important,
because it affects both cooling rates in moderately sized halos and
galaxy colours through the population synthesis models.  Our treatment
of metal enrichment and transport is close to an earlier semi-analytic
model\cite{DeLucia2004}. In it, metals produced and released by
massive stars are placed first into the cold star forming gas, from
which they can be transported into the diffuse hot halo or into the
intergalactic medium by supernova feedback. We assume a homogenous
metallicity (i.e.~perfect mixing) within each of the gas components,
although the hot and cold gas components can have different
metallicities.

\paragraph*{Active galactic nuclei.}

Supermassive black holes are believed to reside at the centre of most,
if not all, spheroidal galaxies, and during their active phases they
power luminous quasars and active galactic nuclei.  There is
substantial observational evidence that suggests a connection between
the formation of galaxies and the build-up of supermassive black holes
(BH). In fact, the energy input provided by BHs may play an important
role in shaping the properties of
galaxies\cite{Silk1998,DiMatteo2005,Springel2005a}, and in reionising
the universe\cite{Haardt1996,Madau2004}.

Our theoretical model for galaxy and AGN formation extends an earlier
semi-analytic model for the joint build-up of the stellar and
supermassive black hole components\cite{Kauffmann2000}.  This adopts
the hypothesis that quasar phases are triggered by galaxy mergers.  In
these events, cold gas is tidally forced into the centre of a galaxy
where it can both fuel a nuclear starburst and be available for
central AGN accretion. We parameterise the efficiency of the feeding
process of the BHs as in the earlier work\cite{Kauffmann2000}, and
normalise it to reproduce the observed scaling relation between the
bulge mass and the BH mass at the present
epoch\cite{Magorrian1998,Ferrarese2000}.  This `quasar mode' of BH
evolution provides the dominant mass growth of the BH population, with
a total cumulative accretion rate that peaks at $z\simeq 3$, similar
to the observed population of quasars.

A new aspect of our model is the addition of a `radio mode' of BH
activity, motivated by the observational phenomenology of nuclear
activity in groups and clusters of galaxies. Here, accretion onto
nuclear supermassive BHs is accompanied by powerful relativistic jets
which can inflate large radio bubbles in clusters, and trigger sound
waves in the intracluster medium (ICM). The buoyant rise of the
bubbles\cite{Churazov2001,Brueggen2002} together with viscous
dissipation of the sound waves\cite{Fabian2003} is capable of
providing a large-scale heating of the ICM, thereby offsetting cooling
losses\cite{Vecchia2004}. These physical processes are arguably the
most likely explanation of the `cooling-flow puzzle': the observed
absence of the high mass dropout rate expected due to the observed
radiative cooling in clusters of galaxies. We parameterise the radio
mode as a mean heating rate into the hot gas proportional to the mass
of the black hole and to the $3/2$ power of the temperature of the hot
gas.  The prefactor is set by requiring a good match to the bright end
of the observed present-day luminosity function of galaxies. The
latter is affected strongly by the radio mode, which reduces the
supply of cold gas to massive central galaxies and thus shuts off
their star formation. Without the radio mode, central cluster galaxies
invariably become too bright and too blue due to excessive cooling
flows. The total BH accretion rate in this radio mode becomes
significant only at very low redshift, but it does not contribute
significantly to the cumulative BH mass density at the present epoch.

\bibliography{si}

\begin{thebibliography}{10}
\expandafter\ifx\csname url\endcsname\relax
  \def\url#1{\texttt{#1}}\fi
\expandafter\ifx\csname urlprefix\endcsname\relax\def\urlprefix{URL }\fi
\providecommand{\bibinfo}[2]{#2}
\providecommand{\eprint}[2][]{\url{#2}}

\bibitem{Bennett2003}
\bibinfo{author}{{Bennett}, C.~L.} \emph{et~al.}
\newblock \bibinfo{title}{{First-Year Wilkinson Microwave Anisotropy Probe
  (WMAP) Observations: Preliminary Maps and Basic Results}}.
\newblock \emph{\bibinfo{journal}{\apjs}} \textbf{\bibinfo{volume}{148}},
  \bibinfo{pages}{1--27} (\bibinfo{year}{2003}).

\bibitem{Spergel2003}
\bibinfo{author}{{Spergel}, D.~N.} \emph{et~al.}
\newblock \bibinfo{title}{{First-Year Wilkinson Microwave Anisotropy Probe
  (WMAP) Observations: Determination of Cosmological Parameters}}.
\newblock \emph{\bibinfo{journal}{\apjs}} \textbf{\bibinfo{volume}{148}},
  \bibinfo{pages}{175--194} (\bibinfo{year}{2003}).

\bibitem{Riess1998}
\bibinfo{author}{{Riess}, A.~G.} \emph{et~al.}
\newblock \bibinfo{title}{{Observational Evidence from Supernovae for an
  Accelerating Universe and a Cosmological Constant}}.
\newblock \emph{\bibinfo{journal}{\aj}} \textbf{\bibinfo{volume}{116}},
  \bibinfo{pages}{1009--1038} (\bibinfo{year}{1998}).

\bibitem{Perlmutter1999}
\bibinfo{author}{{Perlmutter}, S.} \emph{et~al.}
\newblock \bibinfo{title}{{Measurements of Omega and Lambda from 42
  High-Redshift Supernovae}}.
\newblock \emph{\bibinfo{journal}{\apj}} \textbf{\bibinfo{volume}{517}},
  \bibinfo{pages}{565--586} (\bibinfo{year}{1999}).

\bibitem{White1993}
\bibinfo{author}{{White}, S.~D.~M.}, \bibinfo{author}{{Navarro}, J.~F.},
  \bibinfo{author}{{Evrard}, A.~E.} \& \bibinfo{author}{{Frenk}, C.~S.}
\newblock \bibinfo{title}{{The Baryon Content of Galaxy Clusters - a Challenge
  to Cosmological Orthodoxy}}.
\newblock \emph{\bibinfo{journal}{\nat}} \textbf{\bibinfo{volume}{366}},
  \bibinfo{pages}{429} (\bibinfo{year}{1993}).

\bibitem{Davis1985}
\bibinfo{author}{{Davis}, M.}, \bibinfo{author}{{Efstathiou}, G.},
  \bibinfo{author}{{Frenk}, C.~S.} \& \bibinfo{author}{{White}, S.~D.~M.}
\newblock \bibinfo{title}{{The evolution of large-scale structure in a universe
  dominated by cold dark matter}}.
\newblock \emph{\bibinfo{journal}{\apj}} \textbf{\bibinfo{volume}{292}},
  \bibinfo{pages}{371--394} (\bibinfo{year}{1985}).

\bibitem{Colberg2000}
\bibinfo{author}{{Colberg}, J.~M.} \emph{et~al.}
\newblock \bibinfo{title}{{Clustering of galaxy clusters in cold dark matter
  universes}}.
\newblock \emph{\bibinfo{journal}{\mnras}} \textbf{\bibinfo{volume}{319}},
  \bibinfo{pages}{209--214} (\bibinfo{year}{2000}).

\bibitem{Evrard2002}
\bibinfo{author}{{Evrard}, A.~E.} \emph{et~al.}
\newblock \bibinfo{title}{{Galaxy Clusters in Hubble Volume Simulations:
  Cosmological Constraints from Sky Survey Populations}}.
\newblock \emph{\bibinfo{journal}{\apj}} \textbf{\bibinfo{volume}{573}},
  \bibinfo{pages}{7--36} (\bibinfo{year}{2002}).

\bibitem{Wambsganss2004}
\bibinfo{author}{{Wambsganss}, J.}, \bibinfo{author}{{Bode}, P.} \&
  \bibinfo{author}{{Ostriker}, J.~P.}
\newblock \bibinfo{title}{{Giant Arc Statistics in Concord with a Concordance
  Lambda Cold Dark Matter Universe}}.
\newblock \emph{\bibinfo{journal}{\apjl}} \textbf{\bibinfo{volume}{606}},
  \bibinfo{pages}{L93--L96} (\bibinfo{year}{2004}).

\bibitem{Bond1996}
\bibinfo{author}{{Bond}, J.~R.}, \bibinfo{author}{{Kofman}, L.} \&
  \bibinfo{author}{{Pogosyan}, D.}
\newblock \bibinfo{title}{{How filaments of galaxies are woven into the cosmic
  web}}.
\newblock \emph{\bibinfo{journal}{\nat}} \textbf{\bibinfo{volume}{380}},
  \bibinfo{pages}{603} (\bibinfo{year}{1996}).

\bibitem{Jenkins2001}
\bibinfo{author}{{Jenkins}, A.} \emph{et~al.}
\newblock \bibinfo{title}{{The mass function of dark matter haloes}}.
\newblock \emph{\bibinfo{journal}{\mnras}} \textbf{\bibinfo{volume}{321}},
  \bibinfo{pages}{372--384} (\bibinfo{year}{2001}).

\bibitem{Reed2003}
\bibinfo{author}{{Reed}, D.} \emph{et~al.}
\newblock \bibinfo{title}{{Evolution of the mass function of dark matter
  haloes}}.
\newblock \emph{\bibinfo{journal}{\mnras}} \textbf{\bibinfo{volume}{346}},
  \bibinfo{pages}{565--572} (\bibinfo{year}{2003}).

\bibitem{Sheth2002}
\bibinfo{author}{{Sheth}, R.~K.} \& \bibinfo{author}{{Tormen}, G.}
\newblock \bibinfo{title}{{An excursion set model of hierarchical clustering:
  ellipsoidal collapse and the moving barrier}}.
\newblock \emph{\bibinfo{journal}{\mnras}} \textbf{\bibinfo{volume}{329}},
  \bibinfo{pages}{61--75} (\bibinfo{year}{2002}).

\bibitem{Press1974}
\bibinfo{author}{{Press}, W.~H.} \& \bibinfo{author}{{Schechter}, P.}
\newblock \bibinfo{title}{{Formation of Galaxies and Clusters of Galaxies by
  Self-Similar Gravitational Condensation}}.
\newblock \emph{\bibinfo{journal}{\apj}} \textbf{\bibinfo{volume}{187}},
  \bibinfo{pages}{425--438} (\bibinfo{year}{1974}).

\bibitem{Efstathiou1988}
\bibinfo{author}{{Efstathiou}, G.} \& \bibinfo{author}{{Rees}, M.~J.}
\newblock \bibinfo{title}{{High-redshift quasars in the Cold Dark Matter
  cosmogony}}.
\newblock \emph{\bibinfo{journal}{\mnras}} \textbf{\bibinfo{volume}{230}},
  \bibinfo{pages}{5P--11P} (\bibinfo{year}{1988}).

\bibitem{Springel2001b}
\bibinfo{author}{{Springel}, V.}, \bibinfo{author}{{White}, S.~D.~M.},
  \bibinfo{author}{{Tormen}, G.} \& \bibinfo{author}{{Kauffmann}, G.}
\newblock \bibinfo{title}{{Populating a cluster of galaxies - I. Results at
  z=0}}.
\newblock \emph{\bibinfo{journal}{\mnras}} \textbf{\bibinfo{volume}{328}},
  \bibinfo{pages}{726--750} (\bibinfo{year}{2001}).

\bibitem{Kauffmann2000}
\bibinfo{author}{{Kauffmann}, G.} \& \bibinfo{author}{{Haehnelt}, M.}
\newblock \bibinfo{title}{{A unified model for the evolution of galaxies and
  quasars}}.
\newblock \emph{\bibinfo{journal}{\mnras}} \textbf{\bibinfo{volume}{311}},
  \bibinfo{pages}{576--588} (\bibinfo{year}{2000}).

\bibitem{WhiteFrenk1991}
\bibinfo{author}{{White}, S.~D.~M.} \& \bibinfo{author}{{Frenk}, C.~S.}
\newblock \bibinfo{title}{{Galaxy formation through hierarchical clustering}}.
\newblock \emph{\bibinfo{journal}{\apj}} \textbf{\bibinfo{volume}{379}},
  \bibinfo{pages}{52--79} (\bibinfo{year}{1991}).

\bibitem{Kauffmann1993}
\bibinfo{author}{{Kauffmann}, G.}, \bibinfo{author}{{White}, S.~D.~M.} \&
  \bibinfo{author}{{Guiderdoni}, B.}
\newblock \bibinfo{title}{{The Formation and Evolution of Galaxies Within
  Merging Dark Matter Haloes}}.
\newblock \emph{\bibinfo{journal}{\mnras}} \textbf{\bibinfo{volume}{264}},
  \bibinfo{pages}{201--218} (\bibinfo{year}{1993}).

\bibitem{Cole1994}
\bibinfo{author}{{Cole}, S.}, \bibinfo{author}{{Aragon-Salamanca}, A.},
  \bibinfo{author}{{Frenk}, C.~S.}, \bibinfo{author}{{Navarro}, J.~F.} \&
  \bibinfo{author}{{Zepf}, S.~E.}
\newblock \bibinfo{title}{{A Recipe for Galaxy Formation}}.
\newblock \emph{\bibinfo{journal}{\mnras}} \textbf{\bibinfo{volume}{271}},
  \bibinfo{pages}{781--806} (\bibinfo{year}{1994}).

\bibitem{Baugh1996}
\bibinfo{author}{{Baugh}, C.~M.}, \bibinfo{author}{{Cole}, S.} \&
  \bibinfo{author}{{Frenk}, C.~S.}
\newblock \bibinfo{title}{{Evolution of the Hubble sequence in hierarchical
  models for galaxy formation}}.
\newblock \emph{\bibinfo{journal}{\mnras}} \textbf{\bibinfo{volume}{283}},
  \bibinfo{pages}{1361--1378} (\bibinfo{year}{1996}).

\bibitem{Sommerville1999}
\bibinfo{author}{{Somerville}, R.~S.} \& \bibinfo{author}{{Primack}, J.~R.}
\newblock \bibinfo{title}{{Semi-analytic modelling of galaxy formation: the
  local Universe}}.
\newblock \emph{\bibinfo{journal}{\mnras}} \textbf{\bibinfo{volume}{310}},
  \bibinfo{pages}{1087--1110} (\bibinfo{year}{1999}).

\bibitem{Kauffmann1999}
\bibinfo{author}{{Kauffmann}, G.}, \bibinfo{author}{{Colberg}, J.~M.},
  \bibinfo{author}{{Diaferio}, A.} \& \bibinfo{author}{{White}, S.~D.~M.}
\newblock \bibinfo{title}{{Clustering of galaxies in a hierarchical universe -
  I. Methods and results at z=0}}.
\newblock \emph{\bibinfo{journal}{\mnras}} \textbf{\bibinfo{volume}{303}},
  \bibinfo{pages}{188--206} (\bibinfo{year}{1999}).

\bibitem{Fan2003}
\bibinfo{author}{{Fan}, X.} \emph{et~al.}
\newblock \bibinfo{title}{{A Survey of z$>$5.7 Quasars in the Sloan Digital Sky
  Survey. II. Discovery of Three Additional Quasars at z$>$6}}.
\newblock \emph{\bibinfo{journal}{\aj}} \textbf{\bibinfo{volume}{125}},
  \bibinfo{pages}{1649--1659} (\bibinfo{year}{2003}).

\bibitem{Fan2004}
\bibinfo{author}{{Fan}, X.} \emph{et~al.}
\newblock \bibinfo{title}{{A Survey of z$>$5.7 Quasars in the Sloan Digital Sky
  Survey. III. Discovery of Five Additional Quasars}}.
\newblock \emph{\bibinfo{journal}{\aj}} \textbf{\bibinfo{volume}{128}},
  \bibinfo{pages}{515--522} (\bibinfo{year}{2004}).

\bibitem{Tremaine2002}
\bibinfo{author}{{Tremaine}, S.} \emph{et~al.}
\newblock \bibinfo{title}{{The Slope of the Black Hole Mass versus Velocity
  Dispersion Correlation}}.
\newblock \emph{\bibinfo{journal}{\apj}} \textbf{\bibinfo{volume}{574}},
  \bibinfo{pages}{740--753} (\bibinfo{year}{2002}).

\bibitem{Merrit2001}
\bibinfo{author}{{Merritt}, D.} \& \bibinfo{author}{{Ferrarese}, L.}
\newblock \bibinfo{title}{{Black hole demographics from the M{$_{\rm
  BH}$}-{$\sigma$} relation}}.
\newblock \emph{\bibinfo{journal}{\mnras}} \textbf{\bibinfo{volume}{320}},
  \bibinfo{pages}{L30--L34} (\bibinfo{year}{2001}).

\bibitem{Hawkins2003}
\bibinfo{author}{{Hawkins}, E.} \emph{et~al.}
\newblock \bibinfo{title}{{The 2dF Galaxy Redshift Survey: correlation
  functions, peculiar velocities and the matter density of the Universe}}.
\newblock \emph{\bibinfo{journal}{\mnras}} \textbf{\bibinfo{volume}{346}},
  \bibinfo{pages}{78--96} (\bibinfo{year}{2003}).

\bibitem{Benson2000}
\bibinfo{author}{{Benson}, A.~J.}, \bibinfo{author}{{Cole}, S.},
  \bibinfo{author}{{Frenk}, C.~S.}, \bibinfo{author}{{Baugh}, C.~M.} \&
  \bibinfo{author}{{Lacey}, C.~G.}
\newblock \bibinfo{title}{{The nature of galaxy bias and clustering}}.
\newblock \emph{\bibinfo{journal}{\mnras}} \textbf{\bibinfo{volume}{311}},
  \bibinfo{pages}{793--808} (\bibinfo{year}{2000}).

\bibitem{Weinberg2004}
\bibinfo{author}{{Weinberg}, D.~H.}, \bibinfo{author}{{Dav{\' e}}, R.},
  \bibinfo{author}{{Katz}, N.} \& \bibinfo{author}{{Hernquist}, L.}
\newblock \bibinfo{title}{{Galaxy Clustering and Galaxy Bias in a
  {$\Lambda$}CDM Universe}}.
\newblock \emph{\bibinfo{journal}{\apj}} \textbf{\bibinfo{volume}{601}},
  \bibinfo{pages}{1--21} (\bibinfo{year}{2004}).

\bibitem{Padilla2003}
\bibinfo{author}{{Padilla}, N.~D.} \& \bibinfo{author}{{Baugh}, C.~M.}
\newblock \bibinfo{title}{{The power spectrum of galaxy clustering in the APM
  Survey}}.
\newblock \emph{\bibinfo{journal}{\mnras}} \textbf{\bibinfo{volume}{343}},
  \bibinfo{pages}{796--812} (\bibinfo{year}{2003}).

\bibitem{Zehavi2004}
\bibinfo{author}{{Zehavi}, I.} \emph{et~al.}
\newblock \bibinfo{title}{{On Departures from a Power Law in the Galaxy
  Correlation Function}}.
\newblock \emph{\bibinfo{journal}{\apj}} \textbf{\bibinfo{volume}{608}},
  \bibinfo{pages}{16--24} (\bibinfo{year}{2004}).

\bibitem{Norberg2001}
\bibinfo{author}{{Norberg}, P.} \emph{et~al.}
\newblock \bibinfo{title}{{The 2dF Galaxy Redshift Survey: luminosity
  dependence of galaxy clustering}}.
\newblock \emph{\bibinfo{journal}{\mnras}} \textbf{\bibinfo{volume}{328}},
  \bibinfo{pages}{64--70} (\bibinfo{year}{2001}).

\bibitem{Zehavi2002}
\bibinfo{author}{{Zehavi}, I.} \emph{et~al.}
\newblock \bibinfo{title}{{Galaxy Clustering in Early Sloan Digital Sky Survey
  Redshift Data}}.
\newblock \emph{\bibinfo{journal}{\apj}} \textbf{\bibinfo{volume}{571}},
  \bibinfo{pages}{172--190} (\bibinfo{year}{2002}).

\bibitem{Madgwick2003}
\bibinfo{author}{{Madgwick}, D.~S.} \emph{et~al.}
\newblock \bibinfo{title}{{The 2dF Galaxy Redshift Survey: galaxy clustering
  per spectral type}}.
\newblock \emph{\bibinfo{journal}{\mnras}} \textbf{\bibinfo{volume}{344}},
  \bibinfo{pages}{847--856} (\bibinfo{year}{2003}).

\bibitem{deBernardis2000}
\bibinfo{author}{{de Bernardis}, P.} \emph{et~al.}
\newblock \bibinfo{title}{{A flat Universe from high-resolution maps of the
  cosmic microwave background radiation}}.
\newblock \emph{\bibinfo{journal}{\nat}} \textbf{\bibinfo{volume}{404}},
  \bibinfo{pages}{955--959} (\bibinfo{year}{2000}).

\bibitem{Mauskopf2000}
\bibinfo{author}{{Mauskopf}, P.~D.} \emph{et~al.}
\newblock \bibinfo{title}{{Measurement of a Peak in the Cosmic Microwave
  Background Power Spectrum from the North American Test Flight of Boomerang}}.
\newblock \emph{\bibinfo{journal}{\apjl}} \textbf{\bibinfo{volume}{536}},
  \bibinfo{pages}{L59--L62} (\bibinfo{year}{2000}).

\bibitem{Blake2003}
\bibinfo{author}{{Blake}, C.} \& \bibinfo{author}{{Glazebrook}, K.}
\newblock \bibinfo{title}{{Probing Dark Energy Using Baryonic Oscillations in
  the Galaxy Power Spectrum as a Cosmological Ruler}}.
\newblock \emph{\bibinfo{journal}{\apj}} \textbf{\bibinfo{volume}{594}},
  \bibinfo{pages}{665--673} (\bibinfo{year}{2003}).

\bibitem{Jenkins1998}
\bibinfo{author}{{Jenkins}, A.} \emph{et~al.}
\newblock \bibinfo{title}{{Evolution of Structure in Cold Dark Matter
  Universes}}.
\newblock \emph{\bibinfo{journal}{\apj}} \textbf{\bibinfo{volume}{499}},
  \bibinfo{pages}{20--40} (\bibinfo{year}{1998}).

\bibitem{Bardeen1986}
\bibinfo{author}{{Bardeen}, J.~M.}, \bibinfo{author}{{Bond}, J.~R.},
  \bibinfo{author}{{Kaiser}, N.} \& \bibinfo{author}{{Szalay}, A.~S.}
\newblock \bibinfo{title}{{The statistics of peaks of Gaussian random fields}}.
\newblock \emph{\bibinfo{journal}{\apj}} \textbf{\bibinfo{volume}{304}},
  \bibinfo{pages}{15--61} (\bibinfo{year}{1986}).

\bibitem{Adelberger1998}
\bibinfo{author}{{Adelberger}, K.~L.} \emph{et~al.}
\newblock \bibinfo{title}{{A Counts-in-Cells Analysis Of Lyman-break Galaxies
  At Redshift Z \~{} 3}}.
\newblock \emph{\bibinfo{journal}{\apj}} \textbf{\bibinfo{volume}{505}},
  \bibinfo{pages}{18--24} (\bibinfo{year}{1998}).

\bibitem{Cole2005}
\bibinfo{author}{Cole, S.} \emph{et~al.}
\newblock \bibinfo{title}{{The 2dF Galaxy Redshift Survey: Power-spectrum
  analysis of the final dataset and cosmological implications}}.
\newblock \emph{\bibinfo{journal}{\mnras}} \bibinfo{pages}{submitted,
  astro--ph/0501174} (\bibinfo{year}{2005}).

\bibitem{Eisenstein2005}
\bibinfo{author}{Eisenstein, D.~J.} \emph{et~al.}
\newblock \bibinfo{title}{{The 2dF Galaxy Redshift Survey: Power-spectrum
  analysis of the final dataset and cosmological implications}}.
\newblock \emph{\bibinfo{journal}{\apj}} \bibinfo{pages}{submitted,
  astro--ph/0501171} (\bibinfo{year}{2005}).

\bibitem{Springel2001}
\bibinfo{author}{{Springel}, V.}, \bibinfo{author}{{Yoshida}, N.} \&
  \bibinfo{author}{{White}, S.~D.~M.}
\newblock \bibinfo{title}{{GADGET: a code for collisionless and gasdynamical
  cosmological simulations}}.
\newblock \emph{\bibinfo{journal}{New Astronomy}} \textbf{\bibinfo{volume}{6}},
  \bibinfo{pages}{79--117} (\bibinfo{year}{2001}).

\bibitem{Xu1995}
\bibinfo{author}{{Xu}, G.}
\newblock \bibinfo{title}{{A New Parallel N-Body Gravity Solver: TPM}}.
\newblock \emph{\bibinfo{journal}{\apjs}} \textbf{\bibinfo{volume}{98}},
  \bibinfo{pages}{355--366} (\bibinfo{year}{1995}).

\bibitem{Barnes1986}
\bibinfo{author}{{Barnes}, J.} \& \bibinfo{author}{{Hut}, P.}
\newblock \bibinfo{title}{{A Hierarchical O(NlogN) Force-Calculation
  Algorithm}}.
\newblock \emph{\bibinfo{journal}{\nat}} \textbf{\bibinfo{volume}{324}},
  \bibinfo{pages}{446--449} (\bibinfo{year}{1986}).

\bibitem{Hockney1981}
\bibinfo{author}{{Hockney}, R.~W.} \& \bibinfo{author}{{Eastwood}, J.~W.}
\newblock \emph{\bibinfo{title}{{Computer Simulation Using Particles}}}
  (\bibinfo{publisher}{New York: McGraw-Hill, 1981}, \bibinfo{year}{1981}).

\bibitem{Colless2001}
\bibinfo{author}{{Colless}, M.} \emph{et~al.}
\newblock \bibinfo{title}{{The 2dF Galaxy Redshift Survey: spectra and
  redshifts}}.
\newblock \emph{\bibinfo{journal}{\mnras}} \textbf{\bibinfo{volume}{328}},
  \bibinfo{pages}{1039--1063} (\bibinfo{year}{2001}).

\bibitem{White1996}
\bibinfo{author}{White, S. D.~M.}
\newblock \bibinfo{title}{Formation and evolution of galaxies: Les houches
  lectures}.
\newblock In \bibinfo{editor}{Schaefer, R.}, \bibinfo{editor}{Silk, J.},
  \bibinfo{editor}{Spiro, M.} \& \bibinfo{editor}{Zinn-Justin, J.} (eds.)
  \emph{\bibinfo{booktitle}{Cosmology and Large-Scale Structure}}
  (\bibinfo{publisher}{Dordrecht: Elsevier, astro-ph/9410043},
  \bibinfo{year}{1996}).

\bibitem{Seljak1996}
\bibinfo{author}{{Seljak}, U.} \& \bibinfo{author}{{Zaldarriaga}, M.}
\newblock \bibinfo{title}{{A Line-of-Sight Integration Approach to Cosmic
  Microwave Background Anisotropies}}.
\newblock \emph{\bibinfo{journal}{\apj}} \textbf{\bibinfo{volume}{469}},
  \bibinfo{pages}{437--444} (\bibinfo{year}{1996}).

\end{thebibliography}


\begin{thebibliography}{10}
\expandafter\ifx\csname url\endcsname\relax
  \def\url#1{\texttt{#1}}\fi
\expandafter\ifx\csname urlprefix\endcsname\relax\def\urlprefix{URL }\fi
\providecommand{\bibinfo}[2]{#2}
\providecommand{\eprint}[2][]{\url{#2}}

\bibitem{Peebles1970}
\bibinfo{author}{{Peebles}, P.~J.~E.}
\newblock \bibinfo{title}{{Structure of the Coma Cluster of Galaxies}}.
\newblock \emph{\bibinfo{journal}{\aj}} \textbf{\bibinfo{volume}{75}},
  \bibinfo{pages}{13--20} (\bibinfo{year}{1970}).

\bibitem{Miyoshi1975}
\bibinfo{author}{{Miyoshi}, K.} \& \bibinfo{author}{{Kihara}, T.}
\newblock \bibinfo{title}{{Development of the correlation of galaxies in an
  expanding universe}}.
\newblock \emph{\bibinfo{journal}{\pasj}} \textbf{\bibinfo{volume}{27}},
  \bibinfo{pages}{333--346} (\bibinfo{year}{1975}).

\bibitem{White1976}
\bibinfo{author}{{White}, S.~D.~M.}
\newblock \bibinfo{title}{{The dynamics of rich clusters of galaxies}}.
\newblock \emph{\bibinfo{journal}{\mnras}} \textbf{\bibinfo{volume}{177}},
  \bibinfo{pages}{717--733} (\bibinfo{year}{1976}).

\bibitem{Aarseth1979}
\bibinfo{author}{{Aarseth}, S.~J.}, \bibinfo{author}{{Turner}, E.~L.} \&
  \bibinfo{author}{{Gott}, J.~R.}
\newblock \bibinfo{title}{{N-body simulations of galaxy clustering. I - Initial
  conditions and galaxy collapse times}}.
\newblock \emph{\bibinfo{journal}{\apj}} \textbf{\bibinfo{volume}{228}},
  \bibinfo{pages}{664--683} (\bibinfo{year}{1979}).

\bibitem{Efstathiou1981}
\bibinfo{author}{{Efstathiou}, G.} \& \bibinfo{author}{{Eastwood}, J.~W.}
\newblock \bibinfo{title}{{On the clustering of particles in an expanding
  universe}}.
\newblock \emph{\bibinfo{journal}{\mnras}} \textbf{\bibinfo{volume}{194}},
  \bibinfo{pages}{503--525} (\bibinfo{year}{1981}).

\bibitem{Davis1985}
\bibinfo{author}{{Davis}, M.}, \bibinfo{author}{{Efstathiou}, G.},
  \bibinfo{author}{{Frenk}, C.~S.} \& \bibinfo{author}{{White}, S.~D.~M.}
\newblock \bibinfo{title}{{The evolution of large-scale structure in a universe
  dominated by cold dark matter}}.
\newblock \emph{\bibinfo{journal}{\apj}} \textbf{\bibinfo{volume}{292}},
  \bibinfo{pages}{371--394} (\bibinfo{year}{1985}).

\bibitem{White1987}
\bibinfo{author}{{White}, S.~D.~M.}, \bibinfo{author}{{Frenk}, C.~S.},
  \bibinfo{author}{{Davis}, M.} \& \bibinfo{author}{{Efstathiou}, G.}
\newblock \bibinfo{title}{{Clusters, filaments, and voids in a universe
  dominated by cold dark matter}}.
\newblock \emph{\bibinfo{journal}{\apj}} \textbf{\bibinfo{volume}{313}},
  \bibinfo{pages}{505--516} (\bibinfo{year}{1987}).

\bibitem{Carlberg1989}
\bibinfo{author}{{Carlberg}, R.~G.} \& \bibinfo{author}{{Couchman}, H.~M.~P.}
\newblock \bibinfo{title}{{Mergers and bias in a cold dark matter cosmology}}.
\newblock \emph{\bibinfo{journal}{\apj}} \textbf{\bibinfo{volume}{340}},
  \bibinfo{pages}{47--68} (\bibinfo{year}{1989}).

\bibitem{Suto1991}
\bibinfo{author}{{Suto}, Y.} \& \bibinfo{author}{{Suginohara}, T.}
\newblock \bibinfo{title}{{Redshift-space correlation functions in the cold
  dark matter scenario}}.
\newblock \emph{\bibinfo{journal}{\apjl}} \textbf{\bibinfo{volume}{370}},
  \bibinfo{pages}{L15--L18} (\bibinfo{year}{1991}).

\bibitem{Warren1992}
\bibinfo{author}{{Warren}, M.~S.}, \bibinfo{author}{{Quinn}, P.~J.},
  \bibinfo{author}{{Salmon}, J.~K.} \& \bibinfo{author}{{Zurek}, W.~H.}
\newblock \bibinfo{title}{{Dark halos formed via dissipationless collapse. I -
  Shapes and alignment of angular momentum}}.
\newblock \emph{\bibinfo{journal}{\apj}} \textbf{\bibinfo{volume}{399}},
  \bibinfo{pages}{405--425} (\bibinfo{year}{1992}).

\bibitem{Gelb1994}
\bibinfo{author}{{Gelb}, J.~M.} \& \bibinfo{author}{{Bertschinger}, E.}
\newblock \bibinfo{title}{{Cold dark matter. 1: The formation of dark halos}}.
\newblock \emph{\bibinfo{journal}{\apj}} \textbf{\bibinfo{volume}{436}},
  \bibinfo{pages}{467--490} (\bibinfo{year}{1994}).

\bibitem{Zurek1994}
\bibinfo{author}{{Zurek}, W.~H.}, \bibinfo{author}{{Quinn}, P.~J.},
  \bibinfo{author}{{Salmon}, J.~K.} \& \bibinfo{author}{{Warren}, M.~S.}
\newblock \bibinfo{title}{{Large-scale structure after COBE: Peculiar
  velocities and correlations of cold dark matter halos}}.
\newblock \emph{\bibinfo{journal}{\apj}} \textbf{\bibinfo{volume}{431}},
  \bibinfo{pages}{559--568} (\bibinfo{year}{1994}).

\bibitem{Jenkins1998}
\bibinfo{author}{{Jenkins}, A.} \emph{et~al.}
\newblock \bibinfo{title}{{Evolution of Structure in Cold Dark Matter
  Universes}}.
\newblock \emph{\bibinfo{journal}{\apj}} \textbf{\bibinfo{volume}{499}},
  \bibinfo{pages}{20--40} (\bibinfo{year}{1998}).

\bibitem{Governato1999}
\bibinfo{author}{{Governato}, F.} \emph{et~al.}
\newblock \bibinfo{title}{{Properties of galaxy clusters: mass and correlation
  functions}}.
\newblock \emph{\bibinfo{journal}{\mnras}} \textbf{\bibinfo{volume}{307}},
  \bibinfo{pages}{949--966} (\bibinfo{year}{1999}).

\bibitem{Bode2001}
\bibinfo{author}{{Bode}, P.}, \bibinfo{author}{{Bahcall}, N.~A.},
  \bibinfo{author}{{Ford}, E.~B.} \& \bibinfo{author}{{Ostriker}, J.~P.}
\newblock \bibinfo{title}{{Evolution of the Cluster Mass Function: GPC{$^{3}$}
  Dark Matter Simulations}}.
\newblock \emph{\bibinfo{journal}{\apj}} \textbf{\bibinfo{volume}{551}},
  \bibinfo{pages}{15--22} (\bibinfo{year}{2001}).

\bibitem{Colberg2000}
\bibinfo{author}{{Colberg}, J.~M.} \emph{et~al.}
\newblock \bibinfo{title}{{Clustering of galaxy clusters in cold dark matter
  universes}}.
\newblock \emph{\bibinfo{journal}{\mnras}} \textbf{\bibinfo{volume}{319}},
  \bibinfo{pages}{209--214} (\bibinfo{year}{2000}).

\bibitem{Wambsganss2004}
\bibinfo{author}{{Wambsganss}, J.}, \bibinfo{author}{{Bode}, P.} \&
  \bibinfo{author}{{Ostriker}, J.~P.}
\newblock \bibinfo{title}{{Giant Arc Statistics in Concord with a Concordance
  Lambda Cold Dark Matter Universe}}.
\newblock \emph{\bibinfo{journal}{\apjl}} \textbf{\bibinfo{volume}{606}},
  \bibinfo{pages}{L93--L96} (\bibinfo{year}{2004}).

\bibitem{Klypin1983}
\bibinfo{author}{{Klypin}, A.~A.} \& \bibinfo{author}{{Shandarin}, S.~F.}
\newblock \bibinfo{title}{{Three-dimensional numerical model of the formation
  of large-scale structure in the Universe}}.
\newblock \emph{\bibinfo{journal}{\mnras}} \textbf{\bibinfo{volume}{204}},
  \bibinfo{pages}{891--907} (\bibinfo{year}{1983}).

\bibitem{White1983}
\bibinfo{author}{{White}, S.~D.~M.}, \bibinfo{author}{{Frenk}, C.~S.} \&
  \bibinfo{author}{{Davis}, M.}
\newblock \bibinfo{title}{{Clustering in a neutrino-dominated universe}}.
\newblock \emph{\bibinfo{journal}{\apjl}} \textbf{\bibinfo{volume}{274}},
  \bibinfo{pages}{L1--L5} (\bibinfo{year}{1983}).

\bibitem{Centrella1983}
\bibinfo{author}{{Centrella}, J.} \& \bibinfo{author}{{Melott}, A.~L.}
\newblock \bibinfo{title}{{Three-dimensional simulation of large-scale
  structure in the universe}}.
\newblock \emph{\bibinfo{journal}{\nat}} \textbf{\bibinfo{volume}{305}},
  \bibinfo{pages}{196--198} (\bibinfo{year}{1983}).

\bibitem{Park1990}
\bibinfo{author}{{Park}, C.}
\newblock \bibinfo{title}{{Large N-body simulations of a universe dominated by
  cold dark matter}}.
\newblock \emph{\bibinfo{journal}{\mnras}} \textbf{\bibinfo{volume}{242}},
  \bibinfo{pages}{59P--61P} (\bibinfo{year}{1990}).

\bibitem{Bertschinger1991}
\bibinfo{author}{{Bertschinger}, E.} \& \bibinfo{author}{{Gelb}, J.~M.}
\newblock \bibinfo{title}{{Cosmological N-body simulations}}.
\newblock \emph{\bibinfo{journal}{Computers in Physics}}
  \textbf{\bibinfo{volume}{5}}, \bibinfo{pages}{164--175}
  (\bibinfo{year}{1991}).

\bibitem{Bardeen1986}
\bibinfo{author}{{Bardeen}, J.~M.}, \bibinfo{author}{{Bond}, J.~R.},
  \bibinfo{author}{{Kaiser}, N.} \& \bibinfo{author}{{Szalay}, A.~S.}
\newblock \bibinfo{title}{{The statistics of peaks of Gaussian random fields}}.
\newblock \emph{\bibinfo{journal}{\apj}} \textbf{\bibinfo{volume}{304}},
  \bibinfo{pages}{15--61} (\bibinfo{year}{1986}).

\bibitem{Seljak1996}
\bibinfo{author}{{Seljak}, U.} \& \bibinfo{author}{{Zaldarriaga}, M.}
\newblock \bibinfo{title}{{A Line-of-Sight Integration Approach to Cosmic
  Microwave Background Anisotropies}}.
\newblock \emph{\bibinfo{journal}{\apj}} \textbf{\bibinfo{volume}{469}},
  \bibinfo{pages}{437--444} (\bibinfo{year}{1996}).

\bibitem{Spergel2003}
\bibinfo{author}{{Spergel}, D.~N.} \emph{et~al.}
\newblock \bibinfo{title}{{First-Year Wilkinson Microwave Anisotropy Probe
  (WMAP) Observations: Determination of Cosmological Parameters}}.
\newblock \emph{\bibinfo{journal}{\apjs}} \textbf{\bibinfo{volume}{148}},
  \bibinfo{pages}{175--194} (\bibinfo{year}{2003}).

\bibitem{Seljak2004}
\bibinfo{author}{Seljak, U.} \emph{et~al.}
\newblock \bibinfo{title}{{Cosmological parameter analysis including SDSS
  Ly-alpha forest and galaxy bias: constraints on the primordial spectrum of
  fluctuations, neutrino mass, and dark energy}}.
\newblock \emph{\bibinfo{journal}{PRD, submitted, {\em astro-ph/0407372}}}
  (\bibinfo{year}{2004}).

\bibitem{White1996}
\bibinfo{author}{White, S. D.~M.}
\newblock \bibinfo{title}{Formation and evolution of galaxies: Les houches
  lectures}.
\newblock In \bibinfo{editor}{Schaefer, R.}, \bibinfo{editor}{Silk, J.},
  \bibinfo{editor}{Spiro, M.} \& \bibinfo{editor}{Zinn-Justin, J.} (eds.)
  \emph{\bibinfo{booktitle}{Cosmology and Large-Scale Structure}}
  (\bibinfo{publisher}{Dordrecht: Elsevier, astro-ph/9410043},
  \bibinfo{year}{1996}).

\bibitem{Hernquist1993}
\bibinfo{author}{{Hernquist}, L.}, \bibinfo{author}{{Hut}, P.} \&
  \bibinfo{author}{{Makino}, J.}
\newblock \bibinfo{title}{{Discreteness Noise versus Force Errors in N-Body
  Simulations}}.
\newblock \emph{\bibinfo{journal}{\apjl}} \textbf{\bibinfo{volume}{402}},
  \bibinfo{pages}{L85--L88} (\bibinfo{year}{1993}).

\bibitem{Springel2001}
\bibinfo{author}{{Springel}, V.}, \bibinfo{author}{{Yoshida}, N.} \&
  \bibinfo{author}{{White}, S.~D.~M.}
\newblock \bibinfo{title}{{GADGET: a code for collisionless and gasdynamical
  cosmological simulations}}.
\newblock \emph{\bibinfo{journal}{New Astronomy}} \textbf{\bibinfo{volume}{6}},
  \bibinfo{pages}{79--117} (\bibinfo{year}{2001}).

\bibitem{Xu1995}
\bibinfo{author}{{Xu}, G.}
\newblock \bibinfo{title}{{A New Parallel N-Body Gravity Solver: TPM}}.
\newblock \emph{\bibinfo{journal}{\apjs}} \textbf{\bibinfo{volume}{98}},
  \bibinfo{pages}{355--366} (\bibinfo{year}{1995}).

\bibitem{Bode2000}
\bibinfo{author}{{Bode}, P.}, \bibinfo{author}{{Ostriker}, J.~P.} \&
  \bibinfo{author}{{Xu}, G.}
\newblock \bibinfo{title}{{The Tree Particle-Mesh N-Body Gravity Solver}}.
\newblock \emph{\bibinfo{journal}{\apjs}} \textbf{\bibinfo{volume}{128}},
  \bibinfo{pages}{561--569} (\bibinfo{year}{2000}).

\bibitem{Bagla2002}
\bibinfo{author}{{Bagla}, J.~S.}
\newblock \bibinfo{title}{{TreePM: A Code for Cosmological N-Body
  Simulations}}.
\newblock \emph{\bibinfo{journal}{Journal of Astrophysics and Astronomy}}
  \textbf{\bibinfo{volume}{23}}, \bibinfo{pages}{185--196}
  (\bibinfo{year}{2002}).

\bibitem{Barnes1986}
\bibinfo{author}{{Barnes}, J.} \& \bibinfo{author}{{Hut}, P.}
\newblock \bibinfo{title}{{A Hierarchical O(NlogN) Force-Calculation
  Algorithm}}.
\newblock \emph{\bibinfo{journal}{\nat}} \textbf{\bibinfo{volume}{324}},
  \bibinfo{pages}{446--449} (\bibinfo{year}{1986}).

\bibitem{Hockney1981}
\bibinfo{author}{{Hockney}, R.~W.} \& \bibinfo{author}{{Eastwood}, J.~W.}
\newblock \emph{\bibinfo{title}{{Computer Simulation Using Particles}}}
  (\bibinfo{publisher}{New York: McGraw-Hill, 1981}, \bibinfo{year}{1981}).

\bibitem{Salmon1994}
\bibinfo{author}{Salmon, J.~K.} \& \bibinfo{author}{Warren, M.~S.}
\newblock \bibinfo{title}{Skeletons from the treecode closet}.
\newblock \emph{\bibinfo{journal}{J. Comp. Phys.}}
  \textbf{\bibinfo{volume}{111}}, \bibinfo{pages}{136} (\bibinfo{year}{1994}).

\bibitem{Duncan1998}
\bibinfo{author}{{Duncan}, M.~J.}, \bibinfo{author}{{Levison}, H.~F.} \&
  \bibinfo{author}{{Lee}, M.~H.}
\newblock \bibinfo{title}{{A Multiple Time Step Symplectic Algorithm for
  Integrating Close Encounters}}.
\newblock \emph{\bibinfo{journal}{\aj}} \textbf{\bibinfo{volume}{116}},
  \bibinfo{pages}{2067--2077} (\bibinfo{year}{1998}).

\bibitem{Ghigna1998}
\bibinfo{author}{{Ghigna}, S.} \emph{et~al.}
\newblock \bibinfo{title}{{Dark matter haloes within clusters}}.
\newblock \emph{\bibinfo{journal}{\mnras}} \textbf{\bibinfo{volume}{300}},
  \bibinfo{pages}{146--162} (\bibinfo{year}{1998}).

\bibitem{Springel2001b}
\bibinfo{author}{{Springel}, V.}, \bibinfo{author}{{White}, S.~D.~M.},
  \bibinfo{author}{{Tormen}, G.} \& \bibinfo{author}{{Kauffmann}, G.}
\newblock \bibinfo{title}{{Populating a cluster of galaxies - I. Results at
  z=0}}.
\newblock \emph{\bibinfo{journal}{\mnras}} \textbf{\bibinfo{volume}{328}},
  \bibinfo{pages}{726--750} (\bibinfo{year}{2001}).

\bibitem{White1991}
\bibinfo{author}{{White}, S.~D.~M.} \& \bibinfo{author}{{Frenk}, C.~S.}
\newblock \bibinfo{title}{{Galaxy formation through hierarchical clustering}}.
\newblock \emph{\bibinfo{journal}{\apj}} \textbf{\bibinfo{volume}{379}},
  \bibinfo{pages}{52--79} (\bibinfo{year}{1991}).

\bibitem{Kauffmann1993}
\bibinfo{author}{{Kauffmann}, G.}, \bibinfo{author}{{White}, S.~D.~M.} \&
  \bibinfo{author}{{Guiderdoni}, B.}
\newblock \bibinfo{title}{{The Formation and Evolution of Galaxies Within
  Merging Dark Matter Haloes}}.
\newblock \emph{\bibinfo{journal}{\mnras}} \textbf{\bibinfo{volume}{264}},
  \bibinfo{pages}{201--218} (\bibinfo{year}{1993}).

\bibitem{Cole1994}
\bibinfo{author}{{Cole}, S.}, \bibinfo{author}{{Aragon-Salamanca}, A.},
  \bibinfo{author}{{Frenk}, C.~S.}, \bibinfo{author}{{Navarro}, J.~F.} \&
  \bibinfo{author}{{Zepf}, S.~E.}
\newblock \bibinfo{title}{{A Recipe for Galaxy Formation}}.
\newblock \emph{\bibinfo{journal}{\mnras}} \textbf{\bibinfo{volume}{271}},
  \bibinfo{pages}{781--806} (\bibinfo{year}{1994}).

\bibitem{Kauffmann1996}
\bibinfo{author}{{Kauffmann}, G.}
\newblock \bibinfo{title}{{The age of elliptical galaxies and bulges in a
  merger model}}.
\newblock \emph{\bibinfo{journal}{\mnras}} \textbf{\bibinfo{volume}{281}},
  \bibinfo{pages}{487--492} (\bibinfo{year}{1996}).

\bibitem{Kauffmann1997}
\bibinfo{author}{{Kauffmann}, G.}, \bibinfo{author}{{Nusser}, A.} \&
  \bibinfo{author}{{Steinmetz}, M.}
\newblock \bibinfo{title}{{Galaxy formation and large-scale bias}}.
\newblock \emph{\bibinfo{journal}{\mnras}} \textbf{\bibinfo{volume}{286}},
  \bibinfo{pages}{795--811} (\bibinfo{year}{1997}).

\bibitem{Baugh1998}
\bibinfo{author}{{Baugh}, C.~M.}, \bibinfo{author}{{Cole}, S.},
  \bibinfo{author}{{Frenk}, C.~S.} \& \bibinfo{author}{{Lacey}, C.~G.}
\newblock \bibinfo{title}{{The Epoch of Galaxy Formation}}.
\newblock \emph{\bibinfo{journal}{\apj}} \textbf{\bibinfo{volume}{498}},
  \bibinfo{pages}{504--521} (\bibinfo{year}{1998}).

\bibitem{Sommerville1999}
\bibinfo{author}{{Somerville}, R.~S.} \& \bibinfo{author}{{Primack}, J.~R.}
\newblock \bibinfo{title}{{Semi-analytic modelling of galaxy formation: the
  local Universe}}.
\newblock \emph{\bibinfo{journal}{\mnras}} \textbf{\bibinfo{volume}{310}},
  \bibinfo{pages}{1087--1110} (\bibinfo{year}{1999}).

\bibitem{Cole2000}
\bibinfo{author}{{Cole}, S.}, \bibinfo{author}{{Lacey}, C.~G.},
  \bibinfo{author}{{Baugh}, C.~M.} \& \bibinfo{author}{{Frenk}, C.~S.}
\newblock \bibinfo{title}{{Hierarchical galaxy formation}}.
\newblock \emph{\bibinfo{journal}{\mnras}} \textbf{\bibinfo{volume}{319}},
  \bibinfo{pages}{168--204} (\bibinfo{year}{2000}).

\bibitem{Benson2002}
\bibinfo{author}{{Benson}, A.~J.}, \bibinfo{author}{{Lacey}, C.~G.},
  \bibinfo{author}{{Baugh}, C.~M.}, \bibinfo{author}{{Cole}, S.} \&
  \bibinfo{author}{{Frenk}, C.~S.}
\newblock \bibinfo{title}{{The effects of photoionization on galaxy formation -
  I. Model and results at z=0}}.
\newblock \emph{\bibinfo{journal}{\mnras}} \textbf{\bibinfo{volume}{333}},
  \bibinfo{pages}{156--176} (\bibinfo{year}{2002}).

\bibitem{Kauffmann1993tree}
\bibinfo{author}{{Kauffmann}, G.} \& \bibinfo{author}{{White}, S.~D.~M.}
\newblock \bibinfo{title}{{The merging history of dark matter haloes in a
  hierarchical universe.}}
\newblock \emph{\bibinfo{journal}{\mnras}} \textbf{\bibinfo{volume}{261}},
  \bibinfo{pages}{921--928} (\bibinfo{year}{1993}).

\bibitem{Somerville1999tree}
\bibinfo{author}{{Somerville}, R.~S.} \& \bibinfo{author}{{Kolatt}, T.~S.}
\newblock \bibinfo{title}{{How to plant a merger tree}}.
\newblock \emph{\bibinfo{journal}{\mnras}} \textbf{\bibinfo{volume}{305}},
  \bibinfo{pages}{1--14} (\bibinfo{year}{1999}).

\bibitem{Kauffmann1999}
\bibinfo{author}{{Kauffmann}, G.}, \bibinfo{author}{{Colberg}, J.~M.},
  \bibinfo{author}{{Diaferio}, A.} \& \bibinfo{author}{{White}, S.~D.~M.}
\newblock \bibinfo{title}{{Clustering of galaxies in a hierarchical universe -
  I. Methods and results at z=0}}.
\newblock \emph{\bibinfo{journal}{\mnras}} \textbf{\bibinfo{volume}{303}},
  \bibinfo{pages}{188--206} (\bibinfo{year}{1999}).

\bibitem{Croton2005}
\bibinfo{author}{Croton, D.~J.}, \bibinfo{author}{{White}, S.~D.~M.},
  \bibinfo{author}{{Springel}, V.} \& \bibinfo{author}{{et al.}}
\newblock \bibinfo{title}{{The many lives of AGN: super-massive black holes,
  cooling flows, galaxy colours and luminosities}}.
\newblock \emph{\bibinfo{journal}{\mnras}} \bibinfo{pages}{in preparation}
  (\bibinfo{year}{2005}).

\bibitem{Gnedin2000}
\bibinfo{author}{{Gnedin}, N.~Y.}
\newblock \bibinfo{title}{{Cosmological Reionization by Stellar Sources}}.
\newblock \emph{\bibinfo{journal}{\apj}} \textbf{\bibinfo{volume}{535}},
  \bibinfo{pages}{530--554} (\bibinfo{year}{2000}).

\bibitem{Kravtsov2004}
\bibinfo{author}{{Kravtsov}, A.~V.}, \bibinfo{author}{{Gnedin}, O.~Y.} \&
  \bibinfo{author}{{Klypin}, A.~A.}
\newblock \bibinfo{title}{{The Tumultuous Lives of Galactic Dwarfs and the
  Missing Satellites Problem}}.
\newblock \emph{\bibinfo{journal}{\apj}} \textbf{\bibinfo{volume}{609}},
  \bibinfo{pages}{482--497} (\bibinfo{year}{2004}).

\bibitem{Yoshida2002}
\bibinfo{author}{{Yoshida}, N.}, \bibinfo{author}{{Stoehr}, F.},
  \bibinfo{author}{{Springel}, V.} \& \bibinfo{author}{{White}, S.~D.~M.}
\newblock \bibinfo{title}{{Gas cooling in simulations of the formation of the
  galaxy population}}.
\newblock \emph{\bibinfo{journal}{\mnras}} \textbf{\bibinfo{volume}{335}},
  \bibinfo{pages}{762--772} (\bibinfo{year}{2002}).

\bibitem{Helly2003}
\bibinfo{author}{{Helly}, J.~C.} \emph{et~al.}
\newblock \bibinfo{title}{{A comparison of gas dynamics in smooth particle
  hydrodynamics and semi-analytic models of galaxy formation}}.
\newblock \emph{\bibinfo{journal}{\mnras}} \textbf{\bibinfo{volume}{338}},
  \bibinfo{pages}{913--925} (\bibinfo{year}{2003}).

\bibitem{Forcado1997}
\bibinfo{author}{{Forcado-Miro}, M.~I.} \& \bibinfo{author}{{White}, S.~D.~M.}
\newblock \bibinfo{title}{{Radiative shocks in galaxy formation. I: Cooling of
  a primordial plasma with no sources of heating.}}
  \bibinfo{pages}{astro--ph/9712204} (\bibinfo{year}{1997}).

\bibitem{Birnboim2003}
\bibinfo{author}{{Birnboim}, Y.} \& \bibinfo{author}{{Dekel}, A.}
\newblock \bibinfo{title}{{Virial shocks in galactic haloes?}}
\newblock \emph{\bibinfo{journal}{\mnras}} \textbf{\bibinfo{volume}{345}},
  \bibinfo{pages}{349--364} (\bibinfo{year}{2003}).

\bibitem{Keres2004}
\bibinfo{author}{Keres, D.}, \bibinfo{author}{Katz, N.},
  \bibinfo{author}{Weinberg, D.~H.} \& \bibinfo{author}{Dave, R.}
\newblock \bibinfo{title}{{How Do Galaxies Get Their Gas?}}
\newblock \emph{\bibinfo{journal}{\mnras}} \bibinfo{pages}{astro--ph/0407095}
  (\bibinfo{year}{2004}).

\bibitem{Mo1998}
\bibinfo{author}{{Mo}, H.~J.}, \bibinfo{author}{{Mao}, S.} \&
  \bibinfo{author}{{White}, S.~D.~M.}
\newblock \bibinfo{title}{{The formation of galactic discs}}.
\newblock \emph{\bibinfo{journal}{\mnras}} \textbf{\bibinfo{volume}{295}},
  \bibinfo{pages}{319--336} (\bibinfo{year}{1998}).

\bibitem{Kauffmann1996b}
\bibinfo{author}{{Kauffmann}, G.}
\newblock \bibinfo{title}{{Disc galaxies at z=0 and at high redshift: an
  explanation of the observed evolution of damped Lyalpha absorption systems}}.
\newblock \emph{\bibinfo{journal}{\mnras}} \textbf{\bibinfo{volume}{281}},
  \bibinfo{pages}{475--486} (\bibinfo{year}{1996}).

\bibitem{Kennicutt1989}
\bibinfo{author}{{Kennicutt}, R.~C.}
\newblock \bibinfo{title}{{The star formation law in galactic disks}}.
\newblock \emph{\bibinfo{journal}{\apj}} \textbf{\bibinfo{volume}{344}},
  \bibinfo{pages}{685--703} (\bibinfo{year}{1989}).

\bibitem{Kennicutt1998}
\bibinfo{author}{{Kennicutt}, R.~C.}
\newblock \bibinfo{title}{{The Global Schmidt Law in Star-forming Galaxies}}.
\newblock \emph{\bibinfo{journal}{\apj}} \textbf{\bibinfo{volume}{498}},
  \bibinfo{pages}{541--552} (\bibinfo{year}{1998}).

\bibitem{Dekel1986}
\bibinfo{author}{{Dekel}, A.} \& \bibinfo{author}{{Silk}, J.}
\newblock \bibinfo{title}{{The origin of dwarf galaxies, cold dark matter, and
  biased galaxy formation}}.
\newblock \emph{\bibinfo{journal}{\apj}} \textbf{\bibinfo{volume}{303}},
  \bibinfo{pages}{39--55} (\bibinfo{year}{1986}).

\bibitem{Martin1999}
\bibinfo{author}{{Martin}, C.~L.}
\newblock \bibinfo{title}{{Properties of Galactic Outflows: Measurements of the
  Feedback from Star Formation}}.
\newblock \emph{\bibinfo{journal}{\apj}} \textbf{\bibinfo{volume}{513}},
  \bibinfo{pages}{156--160} (\bibinfo{year}{1999}).

\bibitem{Simien1986}
\bibinfo{author}{{Simien}, F.} \& \bibinfo{author}{{de Vaucouleurs}, G.}
\newblock \bibinfo{title}{{Systematics of bulge-to-disk ratios}}.
\newblock \emph{\bibinfo{journal}{\apj}} \textbf{\bibinfo{volume}{302}},
  \bibinfo{pages}{564--578} (\bibinfo{year}{1986}).

\bibitem{Mihos1994}
\bibinfo{author}{{Mihos}, J.~C.} \& \bibinfo{author}{{Hernquist}, L.}
\newblock \bibinfo{title}{{Triggering of starbursts in galaxies by minor
  mergers}}.
\newblock \emph{\bibinfo{journal}{\apjl}} \textbf{\bibinfo{volume}{425}},
  \bibinfo{pages}{L13--L16} (\bibinfo{year}{1994}).

\bibitem{Mihos1996}
\bibinfo{author}{{Mihos}, J.~C.} \& \bibinfo{author}{{Hernquist}, L.}
\newblock \bibinfo{title}{{Gasdynamics and Starbursts in Major Mergers}}.
\newblock \emph{\bibinfo{journal}{\apj}} \textbf{\bibinfo{volume}{464}},
  \bibinfo{pages}{641--663} (\bibinfo{year}{1996}).

\bibitem{Cox2004}
\bibinfo{author}{{Cox}, T.~J.}, \bibinfo{author}{{Primack}, J.},
  \bibinfo{author}{{Jonsson}, P.} \& \bibinfo{author}{{Somerville}, R.~S.}
\newblock \bibinfo{title}{{Generating Hot Gas in Simulations of Disk-Galaxy
  Major Mergers}}.
\newblock \emph{\bibinfo{journal}{\apjl}} \textbf{\bibinfo{volume}{607}},
  \bibinfo{pages}{L87--L90} (\bibinfo{year}{2004}).

\bibitem{Bruzual1993}
\bibinfo{author}{{Bruzual A.}, G.} \& \bibinfo{author}{{Charlot}, S.}
\newblock \bibinfo{title}{{Spectral evolution of stellar populations using
  isochrone synthesis}}.
\newblock \emph{\bibinfo{journal}{\apj}} \textbf{\bibinfo{volume}{405}},
  \bibinfo{pages}{538--553} (\bibinfo{year}{1993}).

\bibitem{Bruzual2003}
\bibinfo{author}{{Bruzual}, G.} \& \bibinfo{author}{{Charlot}, S.}
\newblock \bibinfo{title}{{Stellar population synthesis at the resolution of
  2003}}.
\newblock \emph{\bibinfo{journal}{\mnras}} \textbf{\bibinfo{volume}{344}},
  \bibinfo{pages}{1000--1028} (\bibinfo{year}{2003}).

\bibitem{DeLucia2004}
\bibinfo{author}{{De Lucia}, G.}, \bibinfo{author}{{Kauffmann}, G.} \&
  \bibinfo{author}{{White}, S.~D.~M.}
\newblock \bibinfo{title}{{Chemical enrichment of the intracluster and
  intergalactic medium in a hierarchical galaxy formation model}}.
\newblock \emph{\bibinfo{journal}{\mnras}} \textbf{\bibinfo{volume}{349}},
  \bibinfo{pages}{1101--1116} (\bibinfo{year}{2004}).

\bibitem{Silk1998}
\bibinfo{author}{{Silk}, J.} \& \bibinfo{author}{{Rees}, M.~J.}
\newblock \bibinfo{title}{{Quasars and galaxy formation}}.
\newblock \emph{\bibinfo{journal}{Astr. \& Astrop.}}
  \textbf{\bibinfo{volume}{331}}, \bibinfo{pages}{L1--L4}
  (\bibinfo{year}{1998}).

\bibitem{DiMatteo2005}
\bibinfo{author}{{Di Matteo}, T.}, \bibinfo{author}{{Springel}, V.} \&
  \bibinfo{author}{{Hernquist}, L.}
\newblock \bibinfo{title}{{Energy input from quasars regulates the growth and
  activity of black holes and their host galaxies}}.
\newblock \emph{\bibinfo{journal}{Nature}} \textbf{\bibinfo{volume}{433}},
  \bibinfo{pages}{604--607} (\bibinfo{year}{2005}).

\bibitem{Springel2005a}
\bibinfo{author}{{Springel}, V.}, \bibinfo{author}{{Di Matteo}, T.} \&
  \bibinfo{author}{{Hernquist}, L.}
\newblock \bibinfo{title}{{Black Holes in Galaxy Mergers: The Formation of Red
  Elliptical Galaxies}}.
\newblock \emph{\bibinfo{journal}{\apjl}} \textbf{\bibinfo{volume}{620}},
  \bibinfo{pages}{L79--L82} (\bibinfo{year}{2005}).

\bibitem{Haardt1996}
\bibinfo{author}{{Haardt}, F.} \& \bibinfo{author}{{Madau}, P.}
\newblock \bibinfo{title}{{Radiative Transfer in a Clumpy Universe. II. The
  Ultraviolet Extragalactic Background}}.
\newblock \emph{\bibinfo{journal}{\apj}} \textbf{\bibinfo{volume}{461}},
  \bibinfo{pages}{20--37} (\bibinfo{year}{1996}).

\bibitem{Madau2004}
\bibinfo{author}{{Madau}, P.}, \bibinfo{author}{{Rees}, M.~J.},
  \bibinfo{author}{{Volonteri}, M.}, \bibinfo{author}{{Haardt}, F.} \&
  \bibinfo{author}{{Oh}, S.~P.}
\newblock \bibinfo{title}{{Early Reionization by Miniquasars}}.
\newblock \emph{\bibinfo{journal}{\apj}} \textbf{\bibinfo{volume}{604}},
  \bibinfo{pages}{484--494} (\bibinfo{year}{2004}).

\bibitem{Kauffmann2000}
\bibinfo{author}{{Kauffmann}, G.} \& \bibinfo{author}{{Haehnelt}, M.}
\newblock \bibinfo{title}{{A unified model for the evolution of galaxies and
  quasars}}.
\newblock \emph{\bibinfo{journal}{\mnras}} \textbf{\bibinfo{volume}{311}},
  \bibinfo{pages}{576--588} (\bibinfo{year}{2000}).

\bibitem{Magorrian1998}
\bibinfo{author}{{Magorrian}, J.} \emph{et~al.}
\newblock \bibinfo{title}{{The Demography of Massive Dark Objects in Galaxy
  Centers}}.
\newblock \emph{\bibinfo{journal}{\aj}} \textbf{\bibinfo{volume}{115}},
  \bibinfo{pages}{2285--2305} (\bibinfo{year}{1998}).

\bibitem{Ferrarese2000}
\bibinfo{author}{{Ferrarese}, L.} \& \bibinfo{author}{{Merritt}, D.}
\newblock \bibinfo{title}{{A Fundamental Relation between Supermassive Black
  Holes and Their Host Galaxies}}.
\newblock \emph{\bibinfo{journal}{\apjl}} \textbf{\bibinfo{volume}{539}},
  \bibinfo{pages}{L9--L12} (\bibinfo{year}{2000}).

\bibitem{Churazov2001}
\bibinfo{author}{{Churazov}, E.}, \bibinfo{author}{{Br{\" u}ggen}, M.},
  \bibinfo{author}{{Kaiser}, C.~R.}, \bibinfo{author}{{B{\" o}hringer}, H.} \&
  \bibinfo{author}{{Forman}, W.}
\newblock \bibinfo{title}{{Evolution of Buoyant Bubbles in M87}}.
\newblock \emph{\bibinfo{journal}{\apj}} \textbf{\bibinfo{volume}{554}},
  \bibinfo{pages}{261--273} (\bibinfo{year}{2001}).

\bibitem{Brueggen2002}
\bibinfo{author}{{Br{\" u}ggen}, M.} \& \bibinfo{author}{{Kaiser}, C.~R.}
\newblock \bibinfo{title}{{Hot bubbles from active galactic nuclei as a heat
  source in cooling-flow clusters}}.
\newblock \emph{\bibinfo{journal}{\nat}} \textbf{\bibinfo{volume}{418}},
  \bibinfo{pages}{301--303} (\bibinfo{year}{2002}).

\bibitem{Fabian2003}
\bibinfo{author}{{Fabian}, A.~C.} \emph{et~al.}
\newblock \bibinfo{title}{{A deep Chandra observation of the Perseus cluster:
  shocks and ripples}}.
\newblock \emph{\bibinfo{journal}{\mnras}} \textbf{\bibinfo{volume}{344}},
  \bibinfo{pages}{L43--L47} (\bibinfo{year}{2003}).

\bibitem{Vecchia2004}
\bibinfo{author}{{Vecchia}, C.~D.} \emph{et~al.}
\newblock \bibinfo{title}{{Quenching cluster cooling flows with recurrent hot
  plasma bubbles.}}
\newblock \emph{\bibinfo{journal}{\mnras}} \textbf{\bibinfo{volume}{355}},
  \bibinfo{pages}{995--1004} (\bibinfo{year}{2004}).

\end{thebibliography}

\end{document}